\documentclass[11pt,a4paper]{article}
\usepackage[margin=2 cm]{geometry}
\usepackage{comment}
\usepackage{amssymb,extarrows,graphicx,subfigure,setspace}
\usepackage{cite}
\usepackage{slashed}
\usepackage{tensor}
\usepackage[toc,page]{appendix}
\usepackage{color}
\usepackage{physics}
\usepackage{hyperref}
\usepackage{dirtytalk}
\usepackage[utf8]{inputenc}
\hypersetup{colorlinks=true, linkcolor=blue, citecolor=red, linktoc=page}
\makeatother

\usepackage{amsmath}
\newmuskip\pFqmuskip

\newcommand*\pFq[6][8]{%
  \begingroup 
  \pFqmuskip=#1mu\relax
  \mathcode`\,=\string"8000
  \begingroup\lccode`\~=`\,
  \lowercase{\endgroup\let~}\pFqcomma
  {}_{#2}F_{#3}{\left[\genfrac..{0pt}{}{#4}{#5};#6\right]}%
  \endgroup
}
\newcommand{\pFqcomma}{\mskip\pFqmuskip}

\usepackage{mathrsfs}  
\usepackage{hyperref}

\newcommand{\be}{\begin{equation}}
\newcommand{\ee}{\end{equation}}
\newcommand{\ba}{\begin{eqnarray}}
\newcommand{\ea}{\end{eqnarray}}

\newcommand{\beq}{\begin{equation}}
\newcommand{\eeq}{\end{equation}}
\newcommand{\beqa}{\begin{eqnarray}}
\newcommand{\eeqa}{\end{eqnarray}}

\numberwithin{equation}{section}
\begin{document}
	\onehalfspacing
	\noindent
	
	\begin{titlepage}
		

		\vspace*{20mm}
		\begin{center}
			
			{\Large {\bf Van der Waals Black Holes: Universality, Quantum Corrections, and Topological Classifications}
			}
			
			\vspace*{15mm}
			\vspace*{1mm}
		{\bf \large Ankit Anand$^\star$\footnote{anand@iitk.ac.in} and Saeed Noori Gashti$^\dag$ \footnote{saeed.noorigashti@stu.umz.ac.ir, sn.gashti@du.ac.ir, saeed.noorigashti70@gmail.com}}
		
		\vskip 0.5cm
		{\it $^\star$Department of Physics, Indian Institute of Technology Kanpur, Kanpur 208016, India. \\
        $^\dag$School of Physics, Damghan University, P. O. Box 3671641167, Damghan, Iran.}

			\vspace{0.2cm}

			\vspace*{1cm}
		\end{center}
		
\begin{abstract}

In this paper, we investigate the universal extremality relation and thermodynamic topology of Van der Waals (VdW) black holes-solutions of Einstein's equations whose thermodynamic behavior closely resembles that of Van der Waals fluids. In the classical case, the black hole entropy obeys the Bekenstein-Hawking area law and satisfies the standard universality relation. We then incorporate quantum corrections using three distinct frameworks: the Generalized Uncertainty Principle (GUP), the Extended Uncertainty Principle (EUP), and Rainbow Gravity. While these corrections modify the entropy law, we find that a generalized form of the universal extremality relation still holds. Next, we explore the thermodynamic topology of VdW black holes, focusing on the distribution of topological charges. Our analysis reveals that variations in the black hole and model parameters lead to significant changes in topological classifications and stability, as quantified by winding numbers. In the GUP-corrected case, topological charge distributions exhibit robustness against parameter variations, suggesting classification stability. For EUP-corrected black holes, we identify two distinct topological classes, with some configurations displaying three non-zero topological charges and others maintaining a total charge of zero, despite changes in individual charge counts. The Rainbow Gravity-corrected scenario shows similar consistency in topological behavior.

\end{abstract}

\hspace{1.5cm}\\


\end{titlepage}

\section{Introduction}\label{Sec: Introduction}

The AdS/CFT correspondence \cite{Maldacena:1997re} has significantly renewed interest in the study of asymptotically Anti-de Sitter (AdS) black holes, primarily due to their utility in modeling strongly coupled thermal field theories residing on the AdS boundary. These black holes are not only of holographic interest, but also possess rich thermodynamic structure in the bulk, exhibiting a variety of phase transitions. A paradigmatic example is the first-order Hawking–Page phase transition between thermal AdS radiation and Schwarzschild-AdS black holes \cite{Hawking:1982dh}. By introducing the conserved charges, the thermodynamics of AdS black holes exhibit behavior strikingly analogous to that of a Van der Waals fluid \cite{Chamblin:1999tk, Cvetic:1999ne, Caldarelli:1999xj, Niu:2011tb}. This analogy becomes more precise in the framework of extended phase space thermodynamics \cite{Kubiznak:2012wp, Gunasekaran:2012dq}, where the cosmological constant $\Lambda$ is promoted to a thermodynamic variable interpreted as pressure and allowed to vary in the first law of black hole thermodynamics
\begin{equation} \label{first}
\delta M = T \, \delta S + V \, \delta P + \cdots,
\end{equation}
where $V$ is the thermodynamic volume conjugate to $P$. This formalism permits the definition of a black hole equation of state $P = P(V, T)$, which may be directly compared with that of a conventional fluid. The Van der Waals fluid is characterized by the two-parameter equation of state
\begin{equation} \label{VdW}
T = \left( P + \frac{a}{v^2} \right)(v - b) \ ,
\end{equation}
where $v = V/N$ denotes the specific volume per degree of freedom, with $N$ representing the number of microscopic constituents. The parameters $a > 0$ and $b > 0$ encode the attractive interactions and finite size of fluid molecules, respectively. It has been demonstrated that the thermodynamics of charged and rotating AdS black holes qualitatively reproduce the features of the Van der Waals fluid, including a first-order small/large black hole phase transition analogous to the liquid/gas transition, terminating at a critical point exhibiting classical mean-field critical exponents \cite{Kubiznak:2012wp, Gunasekaran:2012dq}. In both systems, the Gibbs free energy reveals a characteristic swallowtail structure, indicative of a first-order phase transition.

The most profound challenges in theoretical physics are the formulation of a consistent theory that unifies general relativity and quantum mechanics. Several theoretical frameworks have been proposed to bridge this gap, including string theory \cite{Amati:1988tn}, spacetime foam models \cite{Amelino-Camelia:1997ieq}, and loop quantum gravity \cite{Amelino-Camelia:1996bln}. A key prediction shared by these approaches is the existence of a fundamental minimum length scale—commonly the Planck length $L_P \sim 1/E_P$, where $E_P$ is the Planck energy—which acts as a natural cutoff beyond which classical concepts of spacetime cease to apply. In certain scenarios, such as the semiclassical regime of loop quantum gravity \cite{Kozameh:2003rm} and in noncommutative spacetime geometries \cite{Carroll:2001ws}, Lorentz symmetry is effectively broken or deformed. This deviation from conventional Lorentz invariance can be formalized using modified dispersion relations (MDRs) \cite{Ali:2014xqa}, wherein the Planck energy is treated as an invariant scale alongside the speed of light. Such developments have led to the emergence of Doubly Special Relativity (DSR) \cite{Amelino-Camelia:2000stu}, a theoretical extension of special relativity that accommodates both relativistic invariants. Notably, DSR frameworks have been employed to interpret puzzling observations in high-energy astrophysics, including anomalies in ultra-high-energy cosmic rays \cite{Magueijo:2001cr} and the unexpected propagation of TeV photons over cosmological distances \cite{Amelino-Camelia:2002kwj}.

In recent years, topological tools have become central to uncovering the deep structure of phase transitions in black hole thermodynamics, moving beyond conventional thermodynamic potentials. A particularly influential development is the application of Duan’s $\phi$-mapping \cite{Duan:1979ucg, Duan:1999tw} whose core insight of Duan’s theory lies in the fact that topological defects, namely, the zeros of the vector field $\boldsymbol{\phi}$, coincide with critical points in the thermodynamic phase structure. These defects give rise to a conserved topological current derived from a generalized Jacobian tensor, which, under regular conditions, simplifies to the determinant of the Jacobian matrix associated with the $\phi$-mapping. The associated topological invariant, often denoted $W$, encapsulates the global phase structure and is computed via a decomposition involving the Hopf index and the Brouwer degree, which quantify the algebraic winding of the vector field around its zeros\cite{BELFI2003319}. A positive topological charge signals the presence of a locally stable black hole configuration, while a negative value is indicative of thermodynamic instability, often associated with bifurcation or metastability in the phase diagram. This topological formalism not only generalizes classical thermodynamic techniques but also provides a bridge between gravitational thermodynamics and global properties of field configurations, enabling a geometric classification of black hole solutions in diverse spacetime backgrounds.

Within this formalism, the thermodynamic phase space is equipped with a generalized free energy landscape, often constructed from fundamental variables such as the black hole mass and temperature. The equilibrium condition is enforced through the periodicity of Euclidean time $\tau$, which must coincide with the inverse Hawking temperature $T^{-1}$ for a physically admissible (on-shell) configuration \cite{a19, a20}. The free energy function is then analyzed via its gradient structure by introducing a vector field $\boldsymbol{\phi}$, whose components correspond to derivatives with respect to horizon parameters or other intensive/extensive thermodynamic quantities. It has been successfully applied to black holes in asymptotically AdS and dS spacetimes, as well as in theories with higher curvature corrections and nontrivial matter content, demonstrating its wide applicability and predictive power. This topological framework offers a powerful method for investigating the stability and phase transitions of black holes, shedding new light on their thermodynamic behavior. 

The generalized free energy is formally defined as \cite{a19,a20}
\begin{equation}\label{T1}
\mathcal{F} = M - \frac{S}{\tau}
\end{equation}
where \( \tau \) represents the Euclidean time period, and its inverse, \( T \), denotes the system's temperature. Ther are several discussion and for different cases can be found in \cite{20a, 21a, 22a, 23a, 24a, 25a, 26a, 27a, 28a, 29a, 31a, 33a, 34a, 35a, 37a, 38', 38a, 38b, 38c, 39a, 40a, 41a, 42a, 43a, 44a, 44c, 44d, 44e, 44f, 44g, 44h, 44i, 44j, 44k, 44l, 44m, AraujoFilho:2025hnf}. The generalized free energy remains on-shell exclusively when \( \tau \) matches the inverse Hawking temperature. A vector \( \phi \) is then constructed with components obtained from partial derivatives,
\begin{equation}\label{T2}
\phi = \left( \frac{\partial \mathcal{F}}{\partial r_h}, -\cot \Theta \csc \Theta \right)
\end{equation}
At specific angles \( \Theta = 0 \) and \( \Theta = \pi \), \( \phi_{\Theta} \) approaches infinity, with the vector pointing outward. The ranges for the horizon radius \( r_h \) and angular coordinate \( \Theta \) extend from 0 to \( \infty \) and from 0 to \( \pi \), respectively. Using Duan’s \( \phi \)-mapping topological current theory, the topological current is defined as,
\begin{equation}\label{T3}
j^\mu = \frac{1}{2\pi} \epsilon^{\mu \nu \rho} \epsilon^{ab} \partial_\nu n^a \partial_\rho n^b \ , \;\;\;\;\;\; \text{where} \;\;\;\;\; n_i = \frac{\phi_i}{\sqrt{\sum_{i} \phi_i^2}} \;\; \ ,  
\end{equation}
where $\mu, \nu, \rho = 0, 1, 2$ and $i= r_h, \theta$. Using the conservation equation, the current \( j^\mu \) remains nonzero exclusively at points satisfying \( \phi = 0 \). The total charge or topological number \( W \) is determined as
\begin{equation}\label{T4}
W = \int_{\Sigma} j^0 d^2x = \sum_{i=1}^{n} \zeta_i \eta_i = \sum_{i=1}^{n} \omega_i
\end{equation}
where \( \zeta_i \) represents the positive Hopf index\cite{BELFI2003319}, counting the number of loops traced by the vector \( \phi^a \) in the \( \phi \)-space near the zero point \( z_i \), \( \eta_i \) is defined as the sign of \( j^0 (\phi/x)_{z_i} \), which takes values of \( \pm \) and, \( \omega_i \) denotes the winding number associated with the \( i \)-th zero point of \( \phi \) within the region \( \Sigma \).

The paper is structured as follows:

\section{Van der Waals black hole}\label{Sec:Van der Waals black hole}

In this section, we briefly revisit the derivation of an asymptotically AdS black hole solution whose thermodynamic characteristics precisely replicate the prescribed fluid equation of state and its thermodynamics. We start with a static, spherically symmetric metric ansatz as
\begin{equation}\label{Metric Ansatz}
ds^2 = -f\, dt^2+\frac{dr^2}{f}+r^2 \, d\Omega^2 \;\;\;\;\;\text{where}\;\;\;\;\; f(r) = \frac{r^2}{l^2}-\frac{2M}{r}-h(r,P) \ . 
\end{equation}
Our aim is to compute the form of the function $h(r, P)$ by assuming the metric \eqref{Metric Ansatz} is a solution of the Einstein field equations with a given energy-momentum source. The metric ansatz \eqref{Metric Ansatz} leads to a relation between the black hole mass $M$ and the horizon radius $r_h$ given by
\begin{equation}
M = \frac{4}{3} \pi r_h^3 P - \frac{1}{2} r_h h(r_h, P) \ ,
\end{equation}
where $h(r, P)$ encodes deviations from the standard Schwarzschild-AdS form. Enforcing the first law of black hole thermodynamics, Eq.~\eqref{first}, the thermodynamic volume $V$ can be obtained using the Eq.~ \eqref{first}. Using the horizon area is $A = 4\pi r_h^2$, one can compute the specific volume and then express it as
\begin{equation}\label{v of vdW}
v = \frac{k}{4\pi r_h^2} \left( \frac{4}{3} \pi r_h^3 - \frac{r_h}{2} \frac{\partial h(r_h, P)}{\partial P} \right) \ .
\end{equation}
The entropy of the black hole is given by the Bekenstein–Hawking relation, and using that, the Hawking temperature can be computed as
\begin{equation} \label{T of vdW}
T  = 2 r_h P - \frac{h(r_h , P)}{4\pi r_h} - \frac{1}{4\pi} \frac{\partial h(r_h, P)}{\partial r_h} \ .
\end{equation}
This relation can now be compared with a given fluid equation of state $T = T(v, P)$, facilitating the thermodynamic identification. To model the Van der Waals fluid, one compares Eq.~\eqref{T of vdW} with the Van der Waals equation of state \eqref{VdW}. This yields the constraint
\begin{equation}
2 r_h P - \frac{h}{4\pi r_h} - \frac{h'}{4\pi} - \left(P + \frac{a}{v^2} \right)(v - b) = 0,
\end{equation}
where $h' = \partial h/\partial r$, and the specific volume $v$ is substituted from Eq.~\eqref{v of vdW}. This equation defines a partial differential equation (PDE) for the metric function $h(r, P)$, whose solution encodes the gravitational dual of the Van der Waals fluid.

As discussed in \cite{Rajagopal:2014ewa}, one adopt the linear ansatz
\begin{equation} \label{ansatzh}
h(r, P) = A(r) - P \, B(r) \ .
\end{equation}
The solution has been computed as 
\begin{equation}
B(r) = 4\pi \, b \, r \;\;\;;\;\;\; A(r) = -2\pi a + \frac{3\pi a b^2}{r(2r + 3b)} + \frac{4\pi a b}{r} \log \left( \frac{2r}{r_0} + \frac{3b}{r_0} \right)
\end{equation}
where $r_0$ is a constant of integration with dimensions of length. For simplicity $r_0 = 2 b$, which gives the final form of the metric function
\begin{align} \label{metf}
f(r) = 2\pi a-\frac{2M}{r}+\frac{r^2}{l^2}\bigg(1+\frac{3}{2}\frac{b}{r}\bigg)-\frac{3\pi a b^2}{r(2r+3b)}-\frac{4\pi ab}{r}\log\bigg(\frac{r}{b}+\frac{3}{2}\bigg) \ .
\end{align}
This solution describes a black hole spacetime whose thermodynamic behavior reproduces that of a Van der Waals fluid.


\subsection{GUP-Corrected vdW-Black Hole}\label{GUPR}

In this subsection, we will briefly review a generic GUP correction approach by using the simplest form of GUP  in the unit $\hbar=1$ \cite{Xiang:2009yq, Maggiore:1993rv} is 
\begin{equation}\label{GUP}
\Delta x\Delta p\geq1 + \alpha \Delta p^{2}\ ,
\end{equation}
where $\alpha$ is a positive constant. To incorporate the effects of the GUP into black hole thermodynamics, one begins by solving the modified uncertainty relation for the momentum uncertainty $\Delta p$. By solving \eqref{GUP} for $\Delta p$ and expanding the resulting expression in a power series for small $\alpha$, one obtains
\begin{equation}
\label{expandedDeltaP}
\Delta p \geq \frac{1}{\Delta x} + \frac{ \alpha}{\Delta x^3} + \mathcal{O}(\alpha^2) \ .
\end{equation}
It is easy to verify the usual uncertainty relation can be obtained by taking $\alpha \rightarrow 0$. This implies a modified uncertainty product as
\begin{equation}\label{effectivePlanck}
\Delta x \Delta p \geq \left(1 + \frac{\alpha}{\Delta x^2} + \mathcal{O}(\alpha^2)\right) \ ,
\end{equation}
where the second term incorporates the quantum gravity corrections. Next, consider the minimal increase in the horizon area $\Delta A$ due to the absorption of a quantum particle by the black hole. This is bounded below by
\begin{equation}
\label{areaChange}
\Delta A \geq \Delta x \Delta p.
\end{equation}
Assuming the position uncertainty of the particle is of the order of the event horizon radius, and by substituting \eqref{effectivePlanck} into \eqref{areaChange}, one obtains
\begin{equation}
\label{areaChange2}
\Delta A \geq  \gamma  \left(1 + \frac{\alpha}{4 r_h^2} + \mathcal{O}(\alpha^2) \right),
\end{equation}
where $\gamma$ is a dimensionless calibration factor to be determined by consistency with the standard limit $\alpha \to 0$. Since the minimal entropy increase upon absorption of one bit of information is $(\Delta S)_{\text{min}} = \ln 2$, the entropy-area relation gives
\begin{equation}
\label{dA/dS}
\frac{dA}{dS} \simeq  \frac{\gamma}{\ln 2} \left(1 + \frac{\alpha}{4 r_h^2} + \mathcal{O}(\alpha^2) \right).
\end{equation}
Using the thermodynamic relation for black hole temperature, the GUP-modified temperature is easily found as
\begin{equation}
\label{GUPTemp}
T = \frac{\kappa \, \gamma}{8 \pi \ln 2} \left(1 + \frac{\alpha}{4 r_h^2} + \mathcal{O}(\alpha^2) \right) \ ,
\end{equation}
where $\kappa = f'(r_h)/2$ is the surface gravity at the horizon. The value of the calibration factor $\gamma$, is chosen in such a way that the modified temperature reduce to the standard Hawking temperature in the classical limit $\alpha \to 0$, and this requirement fixes $\gamma = 4 \ln 2$, and the final expression for the GUP-corrected black hole temperature becomes
\begin{equation}
\label{GUPTemp2}
T =  \frac{ \kappa}{2\pi} \left(1 + \frac{\alpha}{4 r_h^2} \right) \ ,
\end{equation}
which applies to static, spherically symmetric black hole spacetimes.

Again, we start with a similar metric ansatz as \eqref{Metric Ansatz}, and the function $h(r, P)$ can be obtained from the GUP-corrected black hole temperature. Now, we assume that the given metric is a solution of the Einstein field equations. The general expression of mass and temperature is already in the above section. We can give GUP-corrected thermodynamics. The temperature of the black hole is  
\begin{equation}\label{temperature}
T = \left(1+\frac{\alpha}{4r_{h}^{2}}\right)\left(2Pr_{h}-\frac{h(r_{h},P)}{4\pi r_{h}}-\frac{1}{4\pi}\frac{\partial h(r_{h},P)}{\partial r_{h}}\right) \ .
\end{equation}
Using the first law of thermodynamics, one can compute the entropy as
\begin{equation}\label{entropy GUP vdW}
S = \pi r_{h}^{2}-\frac{\alpha\pi}{4}\ln\left(\frac{4r_{h}^{2}+\alpha}{\alpha}\right) \ ,
\end{equation}
where we choose integration constant as $\frac{\alpha\pi}{4}\ln(\alpha)$ to make a dimensionless logarithmic term in (\ref{entropy GUP vdW}). The specific volume of the black hole is the same as \eqref{v of vdW}. Again, one has to compute the form of $h(r, P)$, using Eq.~\eqref{VdW}, we have to solve  
\begin{equation}
\label{equation}
\left(1+\frac{\alpha}{4r_{h}^{2}}\right)\left(2Pr_{h}-\frac{h}{4\pi r_{h}}-\frac{h'}{4\pi}\right)=\left(P+\frac{a}{v^{2}}\right)(v-b)\,
\end{equation}
where $v=2r_{h}+\frac{3B}{4\pi r_{h}}$. Finally, the form of $B(r)$ and $A(r)$ are  
\begin{eqnarray}
\label{Br2}
B(r) &=& 4\pi br+\left(1+\frac{2b}{3r}\right)\pi\alpha  \;\;\;\;\;;\;\;\;\;
A(r) = \pi  a \left(\frac{3 b^2}{3 b r+2 r^2}+\frac{4 b \log \left(\frac{r}{b}+\frac{3}{2}\right)}{r}-2\right) \nonumber \\
&& \;\;\;\;\;\;\;\;\;\;\;+\frac{\pi  a \alpha }{27 r} \left(\frac{4 \left(\log \left(\frac{r^2}{b^2}\right)-2 \log \left(\frac{r}{b}+\frac{3}{2}\right)\right)}{b}-\frac{3 (36 b+29 r)}{(3 b+2 r)^2}\right) + \mathcal{O}(\alpha^2)\nonumber
\end{eqnarray}
where we choose the suitable integration constant to obtain the dimensionless logarithmic terms.

Finally, putting all those terms together, the form of the metric function is 
\begin{eqnarray}
    && f(r) = 2\pi a-\frac{2M}{r}+\frac{r^2}{l^2}\bigg(1+\frac{3}{2}\frac{b}{r}\bigg)-\frac{3\pi a b^2}{r(2r+3b)}-\frac{4\pi ab}{r}\log\bigg(\frac{r}{b}+\frac{3}{2}\bigg) \nonumber \\
    && +\frac{\alpha}{216 r}  \left(-\frac{32 \pi  a \log \left(\frac{r^2}{b^2}\right)}{b}+\frac{24 \pi  a (36 b+29 r)}{(3 b+2 r)^2}+\frac{64 \pi  a \log \left(\frac{r}{b}+\frac{3}{2}\right)}{b}+\frac{54 b+81 r}{l^2}\right)
\end{eqnarray}
This solution describes a black hole spacetime corrected by GUP, and its thermodynamic behaviour reproduces that of a Van der Waals fluid with GUP correction. It is easy to see that the limit $\alpha \rightarrow 0$ reduces to \eqref{metf}, i.e., the metric of the vdW black hole without modification.
\subsection{EUP-corrected vdW-Black holes}\label{EUP}

In this subsection, we adopt the most elementary form of the Extended Uncertainty Principle (EUP) in the unit $\hbar = 1$~\cite{Kempf:1993bq, Bolen:2004sq, Park:2007az, Mignemi:2009ji} have a form
\begin{equation}
\Delta x \Delta p \geq \left( 1 + \beta (\Delta x)^2 \right).
\end{equation}
Here $\beta$ is the EUP deformation parameter. Within the semi-classical formulation of black hole thermodynamics, the Hawking temperature is given by
\begin{equation}
T = \frac{\kappa}{8\pi} \frac{dA}{dS} \ ,
\end{equation}
where $\kappa$ denotes the surface gravity at the event horizon, while $A$ and $S$ represent the black hole’s surface area and entropy, respectively. According to a heuristic argument based on quantum absorption processes, the minimum change in surface area due to particle absorption is approximated by \cite{Xiang:2009yq}
\begin{equation}
\Delta A \simeq \Delta x \Delta p \ .
\end{equation}
Assuming that the uncertainty in position is of the order of the horizon radius, the minimum area increment within the EUP framework can be written as
\begin{equation}
\Delta A \simeq \frac{\gamma  \left( 1 + \beta r_h^2 \right)}{2} \ ,
\end{equation}
where $\gamma$ is a calibration constant ensuring consistency with the standard Heisenberg Uncertainty Principle in the limit $\beta \to 0$, similar to the GUP case. Again, using the GUP case for the minimum entropy increase associated with a single bit of information as $\left( \Delta S \right)_{\min} = \ln 2$, the area-entropy relation becomes
\begin{equation}
\frac{dA}{dS} \simeq  \frac{\gamma \left( 1 + \beta r_h^2 \right) }{2 \ln 2} \ .
\end{equation}
Fixing the calibration factor as $\gamma = 4 \ln 2$ in the $\beta \to 0$ limit yields the EUP-modified expression for the Hawking temperature
\begin{equation}
T = \frac{ \kappa}{4\pi} \left( 1 + \beta r_h^2 \right) \ . \label{for}
\end{equation}

Now, using the similar metric ansatz as \eqref{Metric Ansatz}, and form of $h(r,P)$ as in \cite{Rajagopal:2014ewa}, the form of $A(r)$ and $B(r)$ are
\begin{eqnarray}
    A(r) &=& \frac{4\pi ab}{r}\log \left(\frac{3}{2}+\frac{r}{b}\right)+\frac{81\beta \pi ab^{5}}{8(3b+2r)^{2}}+\frac{3\pi ab^{2}\left(8-45\beta b^{2}\right) }{8r\left( 3b+2r\right) }-2\pi a\left( 1+\frac{\beta r^{2}}{6}-\frac{\beta br}{2}+\frac{9\beta b^{2}}{8}\right) \nonumber \\
    B(r) &=& 4\pi br-8\pi \beta r^{3}\left( b+\frac{r}{2}\right) \ . 
\end{eqnarray}

Substituting all the terms together the form of metric function is 
\begin{eqnarray}\label{METFUN}
f\left( r\right) &=&\left(-\frac{3 \pi  a b^2}{r (3 b+2 r)}-\frac{4 \pi  a b \log \left(\frac{r}{b}+\frac{3}{2}\right)}{r}+2 \pi  a+\frac{3 b r}{2 l^2}+\frac{r^2}{l^2}-\frac{2 M}{r}\right) \nonumber \\
&& +\beta  \left(-\frac{81 \pi  a b^5}{8 (3 b+2 r)^2}+\frac{135 \pi  a b^4}{8 r (3 b+2 r)}+\frac{9}{4} \pi  a b^2-\pi  a b r+\frac{1}{3} \pi  a r^2-\frac{3 r^3 (2 b+r)}{2 l^2}\right) \ . 
\end{eqnarray}%
This function represents the Van der Waals black hole within the EUP framework, accurate to first order in the deformation parameter $\beta$. In the limit $\beta \rightarrow 0$, the metric reduces to equation \eqref{metf}, recovering the unmodified Van der Waals black hole solution. Equating this to zero, one can easily verify the mass of the black hole as 
\begin{eqnarray}
M &=&-\frac{81}{16}\frac{a\beta b^{5}}{(3b+2r_{h})}\pi r_{h}+\frac{9\pi }{8}%
a\beta b^{2}r_{h}-\frac{3\pi ab^{2}\left( 8-45\beta b^{2}\right) }{%
16(3b+2r_{h})}-\frac{\pi b}{2}r_{h}^{2}(a\beta -4P)  \notag \\
&&-2\pi ab\log \left( \frac{r_{h}}{b}+\frac{3}{2}\right) +\frac{4\pi }{3}%
Pr_{h}^{3}+\pi ar_{h}+\frac{1}{6}\pi a\beta r_{h}^{3}-4\pi \beta
Pr_{h}^{4}\left( b+\frac{r_{h}}{2}\right) \ .  \label{MA}
\end{eqnarray}%
Also the temperature, i.e., the EUP-corrected Hawking temperature of the black hole can be computed as 
\begin{eqnarray}
T &=&\frac{a}{2r_{h}}+2r_{h}P+\beta \frac{ar_{h}}{2}+\frac{ab}{(3b+2r_{h})}%
\left[ \frac{3\beta b}{4}-\frac{2}{r_{h}}-2\beta r_{h}-\frac{81}{32}\frac{%
\beta b^{4}}{r_{h}}\right]  \notag \\
&+&\frac{ab^{2}}{(3b+2r_{h})^{2}}\left[ \frac{81}{16}\beta b^{3}+\frac{3}{2}%
\beta r_{h}+\frac{3}{16r_{h}}\left( 8-45b^{2}\beta \right) \right]  \notag \\
&+&\frac{9}{16}\frac{a\beta b^{2}}{r_{h}}-\frac{b}{2}(a\beta -4P)+\frac{%
a\beta }{4}r_{h}-6\beta r_{h}^{2}P\left( b+\frac{r_{h}}{2}\right) \ .
\label{TH}
\end{eqnarray}
Finally, using them we can compute the expression of entropy as 
\begin{equation}\label{Entropy EUP}
S=\frac{\pi }{\beta }\log \left( 1+\beta r_{h}^{2}\right) \ .
\end{equation}
This reveals that the entropy function exhibits a suppressed growth rate due to the EUP-induced correction term. This decelerative behavior aligns with analogous results derived within the framework of the Generalized Uncertainty Principle.
\subsection{Rainbow Gravity-corrected vdW-Black holes}\label{sec2}
In this subsection, we examine a generic modification to the semi-classical Hawking temperature induced by the rainbow gravity framework. To incorporate rainbow gravity effects by following \cite{Ali:2014xqa}, the spacetime metric is deformed through the introduction of two arbitrary rainbow functions that depend on the ratio of the probe particle energy to the Planck energy. Specifically, the temporal and spatial differentials transform as
\begin{equation}
dt \rightarrow \frac{dt}{\mathcal{F}\left( \frac{E}{E_P} \right)} \;\;\;\;\;;\;\;\;\;\; \qquad dx_i \rightarrow \frac{dx_i}{\mathcal{G}\left( \frac{E}{E_P} \right)} \ .
\end{equation}
Applying these deformations to the metric ansatz \eqref{Metric Ansatz} yields the energy-dependent line element as 
\begin{equation}
ds^{2} = -\frac{f(r)}{\mathcal{F}^{2}\left( \frac{E}{E_P} \right)} dt^2 + \frac{1}{\mathcal{G}^{2}\left( \frac{E}{E_P} \right) f(r)} dr^2 + \frac{r^2}{\mathcal{G}^{2}\left( \frac{E}{E_P} \right)} d\Omega^2 \ . \label{RM}
\end{equation}
The functions $\mathcal{F}$ and $\mathcal{G}$ must reduce to unity in the infrared (low-energy) limit:
\begin{equation}
\lim_{E/E_P \rightarrow 0} \mathcal{F}\left( \frac{E}{E_P} \right) = \lim_{E/E_P \rightarrow 0} \mathcal{G}\left( \frac{E}{E_P} \right) = 1.
\end{equation}
Based on the modified metric \eqref{RM}, the Hawking temperature corrected by rainbow gravity is obtained using the standard surface gravity definition, yielding
\begin{equation}
T_H = \frac{f'(r)|_{r=r_h}}{4\pi} \frac{\mathcal{G}\left( \frac{E}{E_P} \right)}{\mathcal{F}\left( \frac{E}{E_P} \right)} \ .\label{RT}
\end{equation}
To eliminate the energy dependence from the above expression, we invoke the conventional Heisenberg uncertainty principle , which is assumed to remain valid within the Rainbow Gravity framework. According to HUP, $\Delta p \gtrsim 1/\Delta x$, leading to a lower bound on the energy of an emitted particle: $E \gtrsim 1/\Delta x$. Near the event horizon, we approximate $\Delta x \sim r_h$, giving $E \simeq r_h^{-1}$. Substituting this into Eq.~\eqref{RT}, the Rainbow Gravity-modified Hawking temperature becomes
\begin{equation}
T_H = \frac{1}{4\pi} f'(r_h) \frac{\mathcal{G}\left( \frac{1}{r_h E_P} \right)}{\mathcal{F}\left( \frac{1}{r_h E_P} \right)}.
\end{equation}

\quad Again, we follow the similar procedure as we did in van-der waal's case and discussed in \cite{Rajagopal:2014ewa} and compute the $A(r)$ and $B(r)$
\begin{eqnarray}
B\left( r\right) &\simeq& 4b\pi r-\frac{2\pi \gamma }{E_{P}^{2}}\left(1+\frac{2b}{3r}\right)  \nonumber \\
 A\left( r\right) &=&-2\pi a+\frac{3\pi ab^{2}}{r\left(
2r+3b\right) }+\frac{29}{9}\frac{\pi a\gamma }{E_{P}^{2}\left(
2r+3b\right) }+\frac{4\pi ab}{r}\ln \left( \frac{3}{2}+\frac{r}{b%
}\right) +\frac{16}{27}\frac{\pi a\gamma }{bE_{P}^{2}r}\ln \left(
2r+3b\right)  \notag \\
&&-\frac{16}{27}\frac{\pi a\gamma }{bE_{P}^{2}r}\ln r-\frac{5}{3}\frac{%
\pi ab\gamma }{E_{P}^{2}r\left( 2r+3b\right) ^{2}}+\mathcal{O}\left( \gamma ^{2}\right). 
\end{eqnarray}
Finally, using them the metric function forRainbow Gravity-corrected metric function is
\begin{eqnarray}
f\left( r\right) &=&\left(-\frac{3 \pi  a b^2}{r (3 b+2 r)}-\frac{4 \pi  a b \log \left(\frac{r}{b}+\frac{3}{2}\right)}{r}+2 \pi  a+\frac{3 b r}{2 l^2}+\frac{r^2}{l^2}-\frac{2 M}{r}\right)  \\
&& +\gamma  \left(-\frac{29 \pi  a}{9 E_P^2 (3 b+2 r)}+\frac{5 \pi  a b}{3 E_P^2 r (3 b+2 r)^2}+\frac{16 \pi  a \log (r)}{27 b E_P^2 r}-\frac{16 \pi  a \log (3 b+2 r)}{27 b E_P^2 r}-\frac{3 \left(\frac{2 b}{3 r}+1\right)}{4 E_P^2 l^2}\right) \ . \nonumber\label{RGlapse}
\end{eqnarray}
The metric smoothly reduces to its unmodified Van der Waals form, as shown in equation \eqref{metf}, in the limit $\gamma \rightarrow 0$. Now, using this, we can compute the Rainbow Gravity-corrected mass of a VdW black hole is
\small
\begin{eqnarray}
M &=&\frac{4}{3}\pi P r_{h}^{3}+\pi ar_{h}-\frac{3}{2}\frac{\pi ab^{2}}{%
\left( 2r_{h}+3b\right) }-\frac{29}{18}\frac{\pi a\gamma r_{h}}{%
E_{P}^{2}\left( 2r_{h}+3b\right) }-2\pi ab\ln \left( \frac{3}{2}+\frac{r_{h}%
}{b}\right) -\frac{8}{27}\frac{\pi a\gamma }{bE_{P}^{2}}\ln \text{ }\left(
2r_{h}+3b\right)  \notag \\
&&+\frac{8}{27}\frac{\pi a\gamma }{bE_{P}^{2}}\ln r_{h}+\frac{5}{6}\frac{\pi
ab\gamma }{E_{P}^{2}\left( 2r_{h}+3b\right) ^{2}}+2\pi P b r_{h}^{2}-\frac{\pi
r_{h}P\gamma }{E_{P}^{2}}\left( 1+\frac{2b}{3r_{h}}\right) .  \label{mass}
\end{eqnarray}\normalsize
Using the metric function we can also compute the {Rainbow Gravity}-corrected temperature function.
\begin{eqnarray}
T &=&\left( 2bP+2r_{h}P+\frac{a}{2r_{h}}\right) \left( 1-\frac{\gamma }{%
2r_{h}^{2} {E_{P}^{2}}}\right) +\frac{4a\gamma }{27br^{2}_{H}{E_{P}^{2}}}-\frac{P\gamma }{2r_{h}{E_{P}^{2}}} 
\notag \\
&&-\frac{a}{4br_{h}(3b+2r_{h})}\left( \frac{32\gamma }{27{E_{P}^{2}}}+\frac{29 \gamma b}{9{E_{P}^{2}}}%
+8b^{2}\left( 1-\frac{\gamma }{2r_{h}^{2}{E_{P}^{2}}}\right) \right)  \notag \\
&&+\frac{a}{4(3b+2r_{h})^{2}}\left( \frac{58\gamma }{9{E_{P}^{2}}}+\frac{6b^{2}}{r_{h}}%
\left( 1-\frac{\gamma }{2r_{h}^{2}{E_{P}^{2}}}\right) \right) -\frac{5ab\gamma }{%
3r(3b+2r_{h})^{3}}.  \label{tem}
\end{eqnarray}
Finally, using them one can derive the we drive the {Rainbow Gravity}-corrected entropy as
\begin{equation}\label{Entropy for rainbow gravity}
    S\simeq \pi r_{h}^{2}+\frac{\gamma }{E_{P}^{2}}\ln r_{h}+\mathcal{O}\left(
\gamma ^{2}\right) \ .
\end{equation}
It is noteworthy that the second term contributes positively to the entropy, indicating an increase that deviates from the behavior observed in alternative quantum gravity-inspired corrections. Specifically, this result contrasts with the entropy corrections obtained in the GUP framework \cite{Okcu:2022iwl}, where quantum gravitational effects typically suppress entropy growth. This discrepancy highlights the distinctive thermodynamic implications of the Rainbow Gravity-inspired corrections in the theory.\\

The GUP emerges from the expectation that spacetime, at very high energies or extremely small scales, cannot remain smooth and continuous as described by classical physics. This principle finds its roots in theories like string theory and quantum gravity. The central insight is that nature imposes a fundamental limit to how finely one can pinpoint the position of a particle—there's a smallest possible length scale, typically close to the Planck length. As a result, attempts to measure positions with extreme precision lead to increased uncertainty in momentum beyond what standard quantum mechanics predicts. 
In black hole physics, this minimal length has intriguing consequences. It implies that Hawking radiation does not continue endlessly as the black hole shrinks. Instead, the evaporation slows down, eventually halting at a residual mass—often referred to as a remnant. Additionally, the black hole’s temperature gets modified, deviating from the standard inverse-radius behavior at small scales.\\
While GUP focuses on the ultra-small regime, the EUP tackles the opposite limit: extremely large distances and low energies. It suggests that there's a fundamental limit to how precisely momentum can be measured. This idea is particularly relevant in spacetimes with large curvature radii, like those modeled by de Sitter or anti-de Sitter geometries, commonly used in cosmological scenarios.
The EUP framework indicates that at cosmological scales, quantum effects behave differently from what conventional models suggest. For instance, it alters how gravitational clustering occurs over vast regions, influencing thresholds for the formation of cosmic structures. When paired with entropy modifications like the Rényi framework, it gives rise to intriguing shifts in gravitational collapse conditions, modifying the mass limit required for structures like stars or galaxies to become unstable and condense under gravity.\\
Rainbow Gravity stems from efforts to preserve the invariance of physical laws across all energy scales by modifying the standard relationship between energy and spacetime geometry. According to this idea, particles of different energies experience slightly different spacetime metrics. Unlike classical physics, where spacetime is a passive, universal stage for all processes, here it becomes an energy-sensitive fabric. 
This energy dependence has far-reaching implications. It suggests that high-energy particles, like those produced near black holes, might follow different trajectories than low-energy ones. It also modifies how quickly black holes evaporate, how stable certain cosmological models remain, and whether exotic structures like wormholes can persist without collapse. In effect, Rainbow Gravity reinterprets the geometry of the universe as something dynamic and responsive to energy, rather than a fixed backdrop.
 GUP introduces quantum corrections in the realm of the extremely small, emphasizing the idea of a shortest length scale.  
EUP introduces complementary corrections at vast distances, where minimal measurable momentum becomes important.  
Rainbow Gravity shifts the paradigm by proposing that spacetime itself changes depending on the energy of what’s moving through it.

\section{Universality Relations in vdW-black holes} 

A universal relation connecting the entropy and extremality bounds of black holes under perturbative corrections to the theory was proposed in \cite{Goon:2019faz}, given by
\begin{eqnarray}\label{GP Relation}
\frac{\partial M_{\text{ext}}(Q,\, \Vec{\epsilon} )}{\partial \epsilon} = \lim_{M \to M_{\text{ext}}} \left( -T \frac{\partial S(M, Q, \Vec{\epsilon} )}{\partial \epsilon}  \right) \Big|_{M, \Vec{Q}} \ ,
\end{eqnarray}
where $\vec{Q}$ and $\epsilon$ denote thermodynamic quantities such as charge, angular momentum, or other parameters, and $\epsilon$ serves as the perturbative parameter. Here, $M_{\text{ext}}$ represents the extremal mass, and $M$ is the mass of the black hole. To further investigate universal relations in thermodynamics, various studies offer valuable perspectives on the impact of perturbative corrections on black hole properties. These insights contribute to a deeper understanding of fundamental thermodynamic principles, potentially leading to novel discoveries in theoretical physics \cite{2000, 2001, 2002, 2003, 2004}. It has been argued recently in \cite{Anand:2025btp}, if the entropy is not the Bekenstein-Hawking entropy, then this relation can be modified as 
\begin{eqnarray}\label{Anand Relation}
\frac{\partial M_{\text{ext}}(Q,\, \Vec{\epsilon} )}{\partial \epsilon} \;\propto\; \lim_{M \to M_{\text{ext}}} \left( -T \frac{\partial S(M, Q, \Vec{\epsilon} )}{\partial \epsilon}  \right) \Big|_{M, \Vec{Q}} \ ,
\end{eqnarray}
The proportionality constant is $\frac{\partial S_{H}^{\text{BH}}}{\partial S}$ where $S_{H}^{\text{BH}} = \pi r_h^2$ denotes the Bekenstein–Hawking entropy evaluated at the horizon radius obtained from the generalized entropy\footnote{The generalized entropy means the entropy is not just $\pi r_h^2$ it is function of $r_h$ i.e., star with \begin{equation*}
    S=f(r_h) \Rightarrow r_h=f^{-1}(S) =g(S) \ .
\end{equation*} Using this horizon radius one can compute $S_{H}^{\text{BH}}$ as 
\begin{equation*}
    S_{H}^{\text{BH}} = \pi (g(S))^2 \ .
\end{equation*}} $r_h$, which is itself determined from the extremality condition. To test the universality relation, we again follow the procedure of \cite{Goon:2019faz}, the cosmological constant is treated as a perturbative parameter, resulting in a modified gravitational action of the form
\begin{equation*}
\Delta I \propto  \int d^4x \sqrt{-g} \,\epsilon \, \Lambda \ ,
\end{equation*}
where $\epsilon \ll 1$ denotes a small dimensionless perturbation parameter. This perturbation induces a correction in the metric function and the thermodynamic quantities as well. In every considered case, we have started with the same metric ansatz \eqref{Metric Ansatz} and then computed the function $h(r, P)$ in every case, and then the form of the metric function as well. In the next subsections, we start with the expression $h(r, P)$ and with the help of that, we will compute the perturbed thermodynamic quantities and verify the universal relation.

For the vdW-black holes, the $h(r, P)$ for the vdW black hole using the Bekenstein-Hawking entropy relation is 
\begin{eqnarray}
    h_{vdW}(S) = \pi  a \left(\frac{3 \pi  b^2}{3 \sqrt{\pi S} b +2 S}-2\right)+\frac{4 \pi ^{3/2} a b}{\sqrt{S}} \log \left(\frac{3 b}{r_0}+\frac{2}{r_0}\sqrt{\frac{S}{\pi}}\right)-\frac{3 b \sqrt{S}}{2 \sqrt{\pi } l^2} \ .
\end{eqnarray}
Using the Bekenstein-Hawking entropy relation and perturbing the action leads to the modification of thermodynamics. The perturbed mass is 
\begin{equation}\label{M for vdW}
    M = \frac{\sqrt{S} \left(-\pi \, l^2 \, h_{vdW}(S)+S \epsilon +S\right)}{2 \pi ^{3/2} l^2} \ .
\end{equation}
By solving this, we have the expression for the perturbation parameter $\epsilon$ as 
\begin{eqnarray}\label{Epsilon for vdW}
    \epsilon = \frac{\pi  l^2 }{S} h_{vdW(S)}+\frac{2 \pi ^{3/2} l^2 M}{S^{3/2}}-1 \ . 
\end{eqnarray} 
The perturbed temperature is 
\begin{eqnarray}\label{T for vdW}
    T = \frac{1}{4 \pi }\left[\frac{3 S (\epsilon +1)-\pi  l^2 \, h_{vdW(S)}}{l^2 \sqrt{\pi S }  }- \frac{\partial \, h_{vdW(S)}}{\partial S}\right] \ .
\end{eqnarray}
Now, since the entropy of the black hole is the Bekenstein-Hawking entropy so, the universality relation \eqref{GP Relation} or the proportional constant in Eq.~\eqref{Anand Relation} should be $1$. Using the relation \eqref{M for vdW}, \eqref{Epsilon for vdW} and \eqref{T for vdW} we have 
\begin{eqnarray}
    \frac{\partial M_{\text{ext}}}{\partial \epsilon} = \lim_{M \to M_{\text{ext}}} \left( -T \frac{\partial S}{\partial \epsilon}  \right) = \frac{S^{3/2}}{2 \pi ^{3/2} l^2} \ .
\end{eqnarray}
The universal relation is verified in the case of the vdW black holes.
\subsection{Universality Relations in GUP corrected vdW- Black holes}

The $h$-function for the GUP corrected vdW black hole is
\begin{eqnarray}\label{h for GUP-VvdW}
    h_{GvdW}(r) &=& \left(\pi  a \left(\frac{3 b^2}{3 b r+2 r^2}+\frac{4 b \log \left(\frac{r}{b}+\frac{3}{2}\right)}{r}-2\right)-\frac{3 b r}{2 l^2}\right) \\
    &&+\frac{\alpha }{216 r} \left[8 \pi  a \left\{\frac{4}{b} \left(\log \left(\frac{r^2}{b^2}\right)-2 \log \left(\frac{r}{b}+\frac{3}{2}\right)\right)-\frac{3 (36 b+29 r)}{(3 b+2 r)^2}\right\}-\frac{27 (2 b+3 r)}{l^2}\right]  \ . \nonumber
\end{eqnarray}
The entropy of the GUP-corrected vdW black hole is in \eqref{entropy GUP vdW}, we will invert this relation to get the horizon radius in terms of entropy as up to first order of $\alpha$ is  
\begin{eqnarray}\label{rh in S for GUP-VvdW}
    r_h = \sqrt{\frac{S}{\pi }} + \frac{\alpha }{8 \sqrt{\pi  S}} \log \left(\frac{\pi  \alpha }{4 S}\right) \ .
\end{eqnarray}

Now, by perturbing the action again, we have to compute the perturbed thermodynamic quantities. Also, we are computing them at horizon, so we fix $r=r_h$ in \eqref{h for GUP-VvdW}, and to present as a function of entropy, one can use the relation \eqref{rh in S for GUP-VvdW}. The perturbed mass is 
\begin{eqnarray}\label{M for GUP vdW}
    M = \frac{ \left[\left(\alpha  \log \left(\frac{\pi  \alpha }{4 S}\right)+8 S\right)^3-64\pi l^2 S \left(\alpha  \log \left(\frac{\pi  \alpha }{4 S}\right)+8 S\right)\, h_{GvdW}(S)\right]}{1024 (\pi S) ^{3/2} l^2 }+\epsilon \frac{  \left(\alpha  \log \left(\frac{\pi  \alpha }{4 S}\right)+8 S\right)^3}{1024 \pi ^{3/2} l^2 S^{3/2}}
\end{eqnarray}
The perturbed temperature is
\begin{eqnarray}\label{T for GUP vdW}
    T = \frac{-8 \sqrt{S \pi } \frac{\partial h_{GvdW}(S)}{\partial S}-\frac{64 \pi  S h_{GvdW}(S)}{\alpha  \log \left(\frac{\pi  \alpha }{4 S}\right)+8 S}+\frac{3 \left(\alpha  \log \left(\frac{\pi  \alpha }{4 S}\right)+8 S\right)}{l^2}}{32 \pi ^{3/2} \sqrt{S}}+\frac{3 \epsilon  \left(\alpha  \log \left(\frac{\pi  \alpha }{4 S}\right)+8 S\right)}{32 \pi ^{3/2} l^2 \sqrt{S}} \ .
\end{eqnarray}
By solving \eqref{M for GUP vdW}, we can compute the perturbation parameter as
\begin{eqnarray}
    \epsilon =  \frac{64 \pi  l^2 S h_{GvdW}(S)}{\left(\alpha  \log \left(\frac{\pi  \alpha }{4 S}\right)+8 S\right)^2}+\frac{1024 \pi ^{3/2} l^2 M S^{3/2}}{\left(\alpha  \log \left(\frac{\pi  \alpha }{4 S}\right)+8 S\right)^3}-1
\end{eqnarray}
Using this, the R.H.S of Eq.~\eqref{GP Relation} is 
\begin{eqnarray}\label{Tds sepsilon for GUP vdW} 
   -T \left(\frac{\partial S}{\partial \epsilon}\right)\Bigg|_{M_{ext}} = \frac{\sqrt{S} \left(\alpha  \log \left(\frac{\pi  \alpha }{4 S}\right)+8 S\right)^2}{16 \pi ^{3/2} l^2 \left(-2 \alpha -\alpha  \log \left(\frac{\pi  \alpha }{4 S}\right)+8 S\right)} \ .
\end{eqnarray}
From Eq.~\eqref{M for GUP vdW}, it is easy to see that the L.H.S. is not equal to the R.H.S., and it is obvious because the entropy is not just the Bekenstein-Hawking entropy. So, the proportionality constant in Eq.~\eqref{Anand Relation} is not $1$. The proportionality constant can be computed using Eq.~\eqref{rh in S for GUP-VvdW} as 
\begin{eqnarray}\label{Proportinality in GUP-vdW}
   \frac{\partial (\pi r_h^2)}{\partial S} = \frac{\left(-2 \alpha -\alpha  \log \left(\frac{\pi  \alpha }{4 S}\right)+8 S\right) \left(\alpha  \log \left(\frac{\pi  \alpha }{4 S}\right)+8 S\right)}{64 S^2} \ .
\end{eqnarray}
Since the entropy is not just the Bekenstein-Hawking entropy, the generalized universality relation should be verified and can be checked using Eq.~\eqref{M for GUP vdW} and Eq.~\eqref{Tds sepsilon for GUP vdW}, which exactly matches Eq.~\eqref{Proportinality in GUP-vdW}, and this exactly matches the result of \cite{Anand:2025btp}
\subsection{Universality Relations in EUP corrected vdW-Black holes}

The $h-$function for the EUP-corrected vdW black hole is 
\begin{eqnarray}
    && h_{EvdW}(S) = \left[\pi  a \left(\frac{3 \pi  b^2}{3 \sqrt{\pi } b \sqrt{S}+2 S}-2\right)+\frac{4 \pi ^{3/2} a b }{\sqrt{S}}\log \left(\frac{\sqrt{S}}{\sqrt{\pi } b}+\frac{3}{2}\right)-\frac{3 b \sqrt{S}}{2 \sqrt{\pi } l^2}\right] \nonumber \\
    &&\;\;\;\; +\frac{\beta}{24 \pi ^2 l^2 \sqrt{S \left(3 \sqrt{\pi } b+2 \sqrt{S}\right)^4}}  \Bigg[\pi ^2 a l^2 \Bigg(-1215 \pi ^{5/2} b^5+81 \pi ^2 (3 b-16) b^4 \sqrt{S} -486 \pi ^{3/2} b^3 S \nonumber \\
    && \;\;\; +72 \pi  b^2 S^{3/2}+96 \sqrt{\pi } b S^2-32 S^{5/2}\Bigg) +9 \left(7 \sqrt{\pi } b+4 \sqrt{S}\right) \left(3 \sqrt{\pi } b S+2 S^{3/2}\right)^2 -24 \pi ^{5/2} a b l^2 S \nonumber \\
    && \;\;\;\;\;\;\;\;\;\;\;\;\;\;\;\; \left\{3 \sqrt{\pi } b+2 \sqrt{S}\right\}^2 \log \left(\frac{\sqrt{S}}{\sqrt{\pi } b}+\frac{3}{2}\right)\Bigg] \ .\nonumber 
\end{eqnarray}
Inverting the Eq.~\eqref{Entropy EUP}, the horizon radius is 
\begin{eqnarray}\label{Horizon Radius EUP}
    r_h = \frac{\beta  S^{3/2}}{4 \pi ^{3/2}}+\sqrt{\frac{S}{\pi }} \ .
\end{eqnarray}
Now, we perturb the action and compute the perturbation parameter using the perturbed mass as 
\begin{eqnarray}\label{Epsilon for EUP vdW}
    \epsilon = \frac{16 \pi ^3 l^2 h_{EvdW}(S)}{S (\beta  S+4 \pi )^2}+\frac{128 \pi ^{9/2} l^2 M}{S^{3/2} (\beta  S+4 \pi )^3}-1 \ .
\end{eqnarray} 
From the perturbed mass, it is easy to verify that the derivative of the extremal mass with respect to $\epsilon$ is 
\begin{eqnarray}\label{dM by d epsilon}
    \frac{\partial M_{ext}}{\partial \epsilon} = \frac{S^{3/2} (\beta  S+4 \pi )^3}{128 \pi ^{9/2} l^2} \ .
\end{eqnarray}
The temperature will also be perturbed due to the action perturbation, and using that temperature and Eq.~\eqref{Epsilon for EUP vdW}, we have the relation of the R.H.S of Eq.~\eqref{GP Relation} as 
\begin{eqnarray}\label{T ds by d epsilon EUP}
    -T \left(\frac{\partial S}{\partial \epsilon}\right) \Bigg|_{M_{ext}} = -\frac{S^{3/2} (\beta  S+4 \pi )^2}{8 \pi ^{5/2} l^2 (3 \beta  S+4 \pi )} \ . 
\end{eqnarray}
Again, from Eq.~\eqref{dM by d epsilon} and Eq.~\eqref{T ds by d epsilon EUP}, the universal relation \eqref{GP Relation} is not satisfied. The proportionality constant can be computed by dividing Eq.~\eqref{dM by d epsilon} and \eqref{T ds by d epsilon EUP} as 
\begin{eqnarray}\label{EUP Generalized 1}
   \frac{\frac{\partial M_{ext}}{\partial \epsilon}}{-T \left(\frac{\partial S}{\partial \epsilon}\right) \big|_{M_{ext}}} = \frac{(\beta  S+4 \pi ) (3 \beta  S+4 \pi )}{16 \pi ^2} \ .
\end{eqnarray}
Finally, this expression can be verified by writing entropy as the Bekenstein-Hawking entropy using horizon radius as in Eq.~\eqref{Horizon Radius EUP} and differentiating w.r.t. $S$, and it is 
\begin{eqnarray}\label{EUP Generalized 2}
    \frac{\partial }{\partial S}\left(\frac{S (\beta  S+4 \pi )^2}{16 \pi ^2}\right) = \frac{(\beta  S+4 \pi ) (3 \beta  S+4 \pi )}{16 \pi ^2} \ .
\end{eqnarray}
From Eq.~\eqref{EUP Generalized 1} and Eq.~\eqref{EUP Generalized 2} verify the generalized universal relation as discussed in \cite{Anand:2025btp}.
\subsection{Universality Relations in Rainbow-gravity corrected vdW-Black holes}

The $h-$function is 
\begin{eqnarray}
    h_{rgvdW} &=& \left(\pi  a \left(\frac{3 b^2}{3 b r+2 r^2}-2\right)+\frac{4 \pi  a b}{r} \log \left(\frac{r}{b}+\frac{3}{2}\right)-\frac{3 b r}{2 l^2}\right) +\frac{\gamma}{108 E_P^2} \Bigg[\frac{54 b}{l^2 r}+\frac{81}{l^2}\nonumber \\
    && + \frac{12 \pi  a \left(3 b (29 r-5)+58 r^2\right)}{r (3 b+2 r)^2}+\frac{64 \pi  a (\log (3 b+2 r)-\log (r))}{b r}\Bigg] \ .
\end{eqnarray}
The entropy expression for the {Rainbow Gravity}-modified Van der Waals (vdW) black hole is given in Eq.~\eqref{Entropy for rainbow gravity}. To facilitate thermodynamic analysis in terms of entropy, we invert this relation and express the horizon radius $r_h$ as a function of entropy $S$, retaining terms up to the first order in the {Rainbow Gravity} parameter $\gamma$
\begin{eqnarray}\label{rh in S for GUP-rgr}
r_h =  \frac{\gamma}{8 E_P^2 \sqrt{\pi  S}}  \log \left(\frac{\pi  \gamma }{2 E_P^2 S^2}\right)+\sqrt{\frac{S}{\pi }} \ .
\end{eqnarray}
Using this horizon radius, we can compute the Bekenstein-Hawking entropy as 
\begin{eqnarray}
    \pi r_h^2 = \frac{1}{64 E_P^4 S}\left(\gamma  \log \left(\frac{\pi  \gamma }{2 E_P^2 S^2}\right)+8 E_P^2 S\right)^2 \ ,
\end{eqnarray}
and differentiating this respect to $S$, we have 
\begin{eqnarray}\label{Generalized 1 Rgr}
    \frac{\partial}{\partial S}\pi r_h^2 = \frac{\left(-4 \gamma -\gamma  \log \left(\frac{\pi  \gamma }{2 E_P^2 S^2}\right)+8 E_P^2 S\right) \left(\gamma  \log \left(\frac{\pi  \gamma }{2 E_P^2 S^2}\right)+8 E_P^2 S\right)}{64 E_P^4 S^2}
\end{eqnarray}
Subsequently, to study the thermodynamics in the presence of perturbations, we compute the corrected quantities by evaluating the perturbed action at the horizon $r = r_h$. Utilizing the entropy-radius relation from Eq.~\eqref{rh in S for GUP-rgr}, we express all quantities explicitly in terms of entropy.

The perturbed black hole mass takes the form
\begin{eqnarray}\label{M for rgr vdW}
M = \frac{\left(\gamma  \log \left(\frac{\pi  \gamma }{2 E_P^2 S^2}\right)+8 E_P^2 S\right) \left((\epsilon +1) \left(\gamma  \log \left(\frac{\pi  \gamma }{2 E_P^2 S^2}\right)+8 E_P^2 S\right)^2-64 \pi  E_P^4 l^2 S h(S)\right)}{1024 \pi ^{3/2} E_P^6 l^2 S^{3/2}}\ .
\end{eqnarray}
The corresponding perturbed Hawking temperature is
\begin{eqnarray}\label{T for rgr vdW}
T = \frac{1}{32 \pi ^{3/2}}\left[-8 \sqrt{\pi } \frac{\partial h(S)}{\partial S}-\frac{64 \pi  E_P^2 \sqrt{S} h(S)}{\gamma  \log \left(\frac{\pi  \gamma }{2 E_P^2 S^2}\right)+8 E_P^2 S}+\frac{3 (\epsilon +1) \left(\gamma  \log \left(\frac{\pi  \gamma }{2 E_P^2 S^2}\right)+8 E_P^2 S\right)}{E_P^2 l^2 \sqrt{S}} \right] \ .
\end{eqnarray}
Solving Eq.~\eqref{M for rgr vdW} enables us to determine the perturbation parameter $\epsilon$ as
\begin{eqnarray}
\epsilon =  \frac{64 \pi  E_P^4 l^2 S h\left(\frac{\gamma  \log \left(\frac{\pi  \gamma }{2 E_P^2 S^2}\right)+8 E_P^2 S}{8 \sqrt{\pi } E_P^2 \sqrt{S}},P\right)}{\left(\gamma  \log \left(\frac{\pi  \gamma }{2 E_P^2 S^2}\right)+8 E_P^2 S\right)^2}+\frac{1024 \pi ^{3/2} E_P^6 l^2 M S^{3/2}}{\left(\gamma  \log \left(\frac{\pi  \gamma }{2 E_P^2 S^2}\right)+8 E_P^2 S\right)^3}-1 \ .
\end{eqnarray}
With this expression, the right-hand side (R.H.S.) of the generalized first law relation, Eq.~\eqref{GP Relation}, becomes
\begin{eqnarray}\label{Tds sepsilon for rgr P vdW}
-T \left( \frac{\partial S}{\partial \epsilon} \right)\Bigg|_{M_{ext}} = \frac{\sqrt{S} \left(\gamma  \log \left(\frac{\pi  \gamma }{2 E_P^2 S^2}\right)+8 E_P^2 S\right)^2}{16 \pi ^{3/2} E_P^2 l^2 \left(-4 \gamma -\gamma  \log \left(\frac{\pi  \gamma }{2 E_P^2 S^2}\right)+8 E_P^2 S\right)} \ .
\end{eqnarray}
It is evident from Eq.~\eqref{M for GUP vdW} that the left-hand side (L.H.S.) of Eq.~\eqref{GP Relation} does not match the R.H.S., which is expected since the entropy is no longer purely Bekenstein-Hawking in form. Consequently, the proportionality constant in the generalized relation, Eq.~\eqref{Anand Relation}, deviates from unity.

Differentiating Eq.~\eqref{M for rgr vdW} w.r.t. $\epsilon$ and dividing by Eq.~\eqref{Tds sepsilon for rgr P vdW}, we have 
\begin{eqnarray}\label{Generalized 2 Rgr}
\frac{\frac{\partial M_{ext}}{\partial \epsilon}}{-T \left(\frac{\partial S}{\partial \epsilon}\right) \big|_{M_{ext}}} = \frac{\left(-4 \gamma -\gamma  \log \left(\frac{\pi  \gamma }{2 E_P^2 S^2}\right)+8 E_P^2 S\right) \left(\gamma  \log \left(\frac{\pi  \gamma }{2 E_P^2 S^2}\right)+8 E_P^2 S\right)}{64 E_P^4 S^2} \ .
\end{eqnarray}
This expression exactly matched with Eq.~\eqref{Generalized 1 Rgr} confirms that the entropy deviates from the classical Bekenstein-Hawking form, and therefore, the generalized universality relation is satisfied \cite{Anand:2025btp}.

\section{Thermodynamic Topology for vdW black holes}

\begin{figure}[]
 \begin{center}
 \subfigure[]{
 \includegraphics[height=5cm,width=7cm]{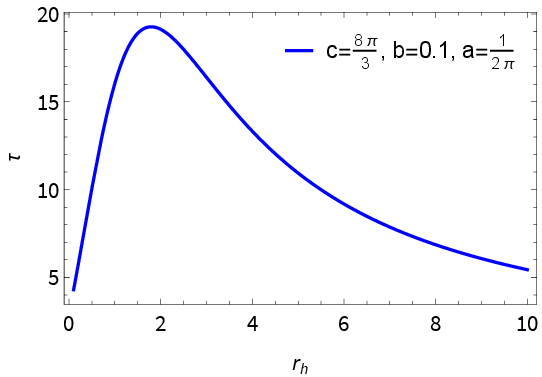}
 \label{100a}}
 \subfigure[]{
 \includegraphics[height=5cm,width=7cm]{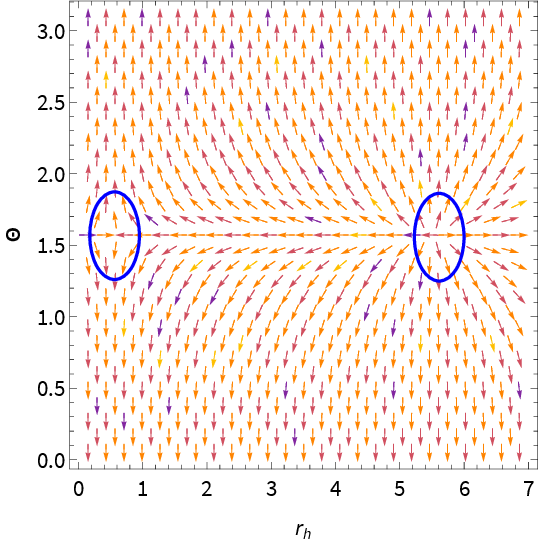}
 \label{100b}}\\
 \subfigure[]{
 \includegraphics[height=5cm,width=7cm]{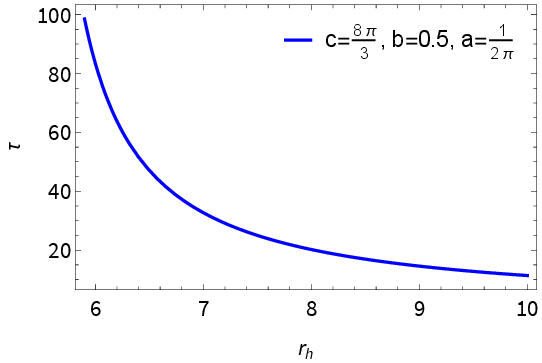}
 \label{100c}}
 \subfigure[]{
 \includegraphics[height=5cm,width=7cm]{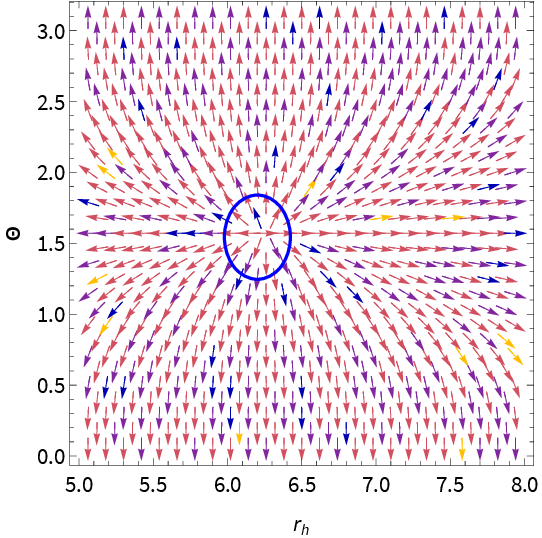}
 \label{100d}}\\
 \subfigure[]{
 \includegraphics[height=5cm,width=7cm]{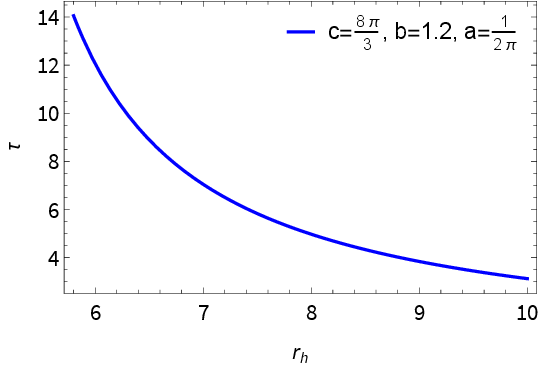}
 \label{100e}}
 \subfigure[]{
 \includegraphics[height=5cm,width=7cm]{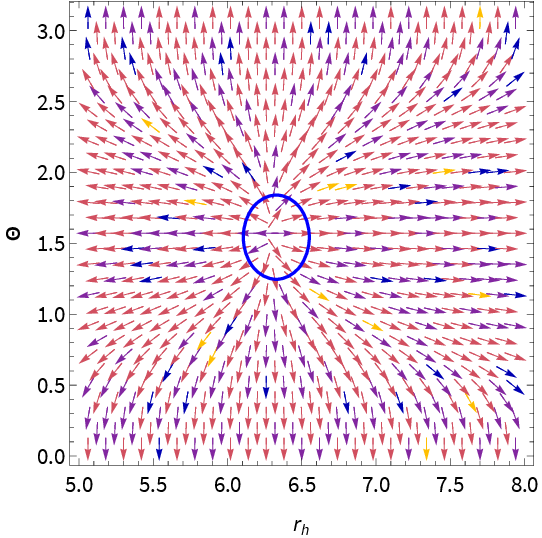}
 \label{100f}}
 \caption{\small{The $(\tau \text{ vs. } r_h)$ diagram, illustrating variations in free parameters for van der Waals (vdW) black holes, is presented in Figs. (\ref{100a}), (\ref{100c}), and (\ref{100e}). Additionally, the normal vector field \( n \) in the \((r_h - \Theta)\) plane is depicted, with Zero Points (ZPs) located at specific coordinates \((r_h, \Theta)\). These ZPs correspond to parameter values \( b = 0.1, 0.5, 1.2 \), along with fixed values of \( c = 8\pi/3 \), \( \lambda = 1 \), and \( a = 1/2\pi \), as shown in Figs. (\ref{100b}), (\ref{100d}), and (\ref{100f}), respectively.}}
 \label{m1}
\end{center}
 \end{figure}

In this section, we discuss the thermodynamic topology and photon sphere for vdW black holes. Thus, based on Eq.~\eqref{T1} for van der Waals (vdW) black holes, the generalized Helmholtz free energy is computed
\begin{equation}\label{T5}
\mathcal{F}=\pi  a \left(r_h-\frac{3 b^2}{6 b+4 r_h}\right)-2 \pi  a b \log \left(\frac{3 b+2 r_h}{r_0}\right)+\frac{r_h^2 \left(-36 \pi  b-9 c r_h+8 \pi  r_h \left(8 \pi  l^2 P+3\right)\right)}{48 \pi  l^2}-\frac{\pi  r_h^2}{\tau } \ ,
\end{equation}
$c = 8\pi/3$ is defined just for computational purposes. Furthermore, from Eq. (\ref{T2}), the components of the vector field $\phi$ can be determined as,
\begin{equation}\label{T6}
\begin{split}
\phi^{r_h}=\frac{4 \pi  a r_h (b+r_h)}{(3 b+2 r_h)^2}+\frac{3 r_h (-8 \pi  b-3 c r_h+8 \pi  r_h)}{16 \pi  l^2}+\frac{2 \pi  r_h (2 P r_h \tau -1)}{\tau } \;\;\;\;\;;\;\;\; \phi ^{\theta }=-\frac{\cot (\theta )}{\sin (\theta )} \ .
\end{split}
\end{equation}
Additionally by solving Eq.~\eqref{T5}, we obtain,
\begin{equation}\label{T7}
\begin{split}
&\tau = 32 \pi ^2 l^2 (3 b+2 r_h)^2 \; \bigg[64 \pi ^2 a b l^2+64 \pi ^2 a l^2 r_h-216 \pi  b^3-9 b^2 r_h \left(9 c+8 \pi  \left(1-8 \pi  l^2 P\right)\right)\\&+12 b r_h^2 \left(16 \pi  \left(4 \pi  l^2 P+1\right)-9 c\right)+4 r_h^3 \left(8 \pi  \left(8 \pi  l^2 P+3\right)-9 c\right)\bigg]^{-1}
\end{split}
\end{equation}
This study explores the thermodynamic topology of van der Waals (vdW) black holes, with illustrations detailing topological charge distributions. The normalized field lines, presented in Fig. (\ref{m1}), depict the topological charge characteristics. Fig.~(\ref{100b}) illustrates the presence of two zero points at specific coordinates \((r_h, \Theta)\), corresponding to parameter values \( b = 0.1 \), \( c = 8\pi/3 \), \( \lambda = 1 \), and \( a = 1/2\pi \). In contrast, Figs. (\ref{100d}) and (\ref{100f}) reveal a single zero point for higher values of \( b = 0.5, 1.2 \). These zero points, representing topological charges, lie within blue contour loops, with their arrangement dictated by variations in the parameter \( b \). A significant feature emerges from these results: when \( b = 0.1 \), two topological charges \( (\omega = +1, -1) \) are present, yielding a total topological charge of \( W = 0 \), as indicated in Fig. (\ref{100b}). Stability analysis of the black hole is conducted through winding number evaluations. As the parameter \( b \) increases, the classification undergoes a transition, as seen in Figs. (\ref{100d}) and (\ref{100f}). In these cases, a single topological charge \( (\omega = +1) \) is observed, corresponding to a total charge of \( W = +1 \).

\subsection{Thermodynamic Topology in GUP corrected vdW- Black holes}

In this subsection, we examine topological charges along with their classification. Based on Eq. (\ref{T1}) for the GUP-corrected black hole, the generalized Helmholtz free energy is expressed as,
\begin{equation}\label{g1}
\begin{split}
&\mathcal{F}=\frac{\pi}{12} \Bigg\{-\frac{6 a \alpha  b (7 \alpha ^2+108 b^4+45 \alpha  b^2) \log (\frac{\alpha +4 r_h^2}{4 b^2})}{(\alpha +9 b^2)^3}-\frac{12 a b (-5 \alpha ^3+1458 b^6+378 \alpha  b^4+9 \alpha ^2 b^2) \log (\frac{r_h}{b}+\frac{3}{2})}{(\alpha +9 b^2)^3}\\&+\frac{6 a b^2 (-243 b^4 (3 b+2 r_h)+9 \alpha  b^2 (18 b+17 r_h)+\alpha ^2 (57 b+43 r_h))}{(\alpha +9 b^2)^2 (3 b+2 r_h)^2}-\frac{3 a \alpha ^{3/2} (5 \alpha ^2-27 b^4+2 \alpha  b^2) \tan ^{-1}(\frac{2 r_h}{\sqrt{\alpha }})}{(\alpha +9 b^2)^3}\\&+\frac{8 r_h (2 b (\alpha +6 r_h^2)+3 \alpha  r_h) (4 r_h^2 (3 b+2 r_h) ( (3 b+2 r_h)^2-2 a (b+r_h))+a \alpha  (b+2 r_h) (2 b+3 r_h))}{(3 b+2 r_h)^3 (\alpha +4 r_h^2) (\alpha +8 r_h (b+r_h))}\\&+12 a r_h+16 P r_h^3+\frac{3 (\alpha  \log (\frac{4 r_h^2}{\alpha }+1)-4 r_h^2)}{\tau }\Bigg\} \ .
\end{split}
\end{equation}
Furthermore, from Eq.~(\ref{T2}), the components of the vector field $\phi$ can be determined as
\begin{eqnarray}\label{g2}
&&\phi^{r_h}=\bigg\{4 \pi  (a \tau  (-4 \alpha ^3 r_h^2 (45 b^4+200 b^3 r_h+419 b^2 r_h^2+448 b r_h^3+168 r_h^4)+256 \alpha  r_h^5 (b+r_h) (-36 b^4-57 b^3 r_h \nonumber \\
&& -4 b^2 r_h^2+20 b r_h^3+12 r_h^4)-16 \alpha ^2 r_h^3 (96 b^5+589 b^4 r_h+1318 b^3 r_h^2+1539 b^2 r_h^3+960 b r_h^4+228 r_h^5)+3072 r_h^8 \nonumber \\
&&(b  +r_h)^2 (3 b+2 r_h) (5 b+2 r_h)+ \alpha^4 b (2 b+3 r_h)(b+4 r_h) (3 b+ 5 r_h))- 124416b^6 r_h^7-31104 \alpha  b^6 r_h^5 \nonumber \\
&&-580608 b^5 r_h^8 -176256 \alpha  b^5 r_h^6-7776 \alpha^2 b^5 r_h^4 -1119744 b^4 r_h^9-393984 \alpha  b^4 r_h^7-30456 \alpha ^2 b^4 r_h^5-486 \alpha ^3 b^4 r_h^3 \nonumber \\
&& -1142784 b^3 r_h^{10}-451584 \alpha  b^3 r_h^8-46656 \alpha ^2 b^3 r_h^6-1296 \alpha ^3 b^3 r_h^4+r_h^2 \tau  (3 b+2 r_h)^4 (64 b^2 r_h (3 P r_h (\alpha +4 r_h^2)^2 \nonumber \\
&& +(\alpha ^2+24 r_h^4+12 \alpha  r_h^2))+4 b (12 P r_h (\alpha +8 r_h^2) (\alpha +4 r_h^2)^2+ (3 \alpha ^3+192 r_h^6+232 \alpha  r_h^4+78 \alpha ^2 r_h^2))+3 (P (\alpha \nonumber \\
&&+4 r_h^2)^2 (\alpha +8 r_h^2)^2+8 \alpha ^2 r_h  (\alpha +6 r_h^2)))-651264 b^2 r_h^{11}-282624 \alpha  b^2 r_h^9-35136 \alpha ^2 b^2 r_h^7-1296 \alpha ^3 b^2 r_h^5 \nonumber \\
&&-196608 b r_h^{12}-92160 \alpha  b r_h^{10}-13056 \alpha ^2 b r_h^8-576 \alpha ^3 b r_h^6-24576 r_h^{13}-12288 \alpha  r_h^{11}-1920 \alpha ^2 r_h^9-96 \alpha ^3 r_h^7)\bigg\} \nonumber \\
&& \times\bigg[3 \tau  (3 b+2 r_h)^4 (\alpha +4 r_h^2)^2 (\alpha +8 r_h (b+r_h))^2\bigg]^{-1} \;\;\;\;\;\;\; \text{and}\;\;\;\;\;\;\;\phi ^{\Theta }=-\frac{\cot (\Theta )}{\sin (\Theta )} \ .
\end{eqnarray}
Additionally, we derive,
\begin{eqnarray}\label{g4}
&&\tau =\bigg(24 r_h^3 (3 b+2 r_h)^4 (4 r_h^2+\alpha ) (8 r_h^2+8 b r_h+\alpha )^2\bigg)\times\bigg[196608 P r_h^{14}+1572864 b P r_h^{13}+49152 a r_h^{12}\nonumber \\ &&+5210112 b^2 P r_h^{12}+49152 b  r_h^{12}+147456 P \alpha  r_h^{12}+294912 a b r_h^{11}+9142272 b^3 P r_h^{11}+393216 b^2  r_h^{11}\nonumber \\ &&+1130496 b P \alpha  r_h^{11}+626688 a b^2 r_h^{10}+39936 P \alpha ^2 r_h^{10}+8957952 b^4 P r_h^{10}+1253376 b^3  r_h^{10}+12288 a \alpha  r_h^{10}\nonumber \\ &&+3563520 b^2 P \alpha  r_h^{10}+59392 b  \alpha  r_h^{10}+565248 a b^3 r_h^9+288768 b P \alpha ^2 r_h^9+9216  \alpha ^2 r_h^9+4644864 b^5 P r_h^9\nonumber \\ &&+1990656 b^4  r_h^9+32768 a b \alpha  r_h^9+5898240 b^3 P \alpha  r_h^9+405504 b^2  \alpha  r_h^9+184320 a b^4 r_h^8+4608 P \alpha ^3 r_h^8\nonumber \\ &&-14592 a \alpha ^2 r_h^8+846336 b^2 P \alpha ^2 r_h^8+75264 b  \alpha ^2 r_h^8+995328 b^6 P r_h^8+1575936 b^5  r_h^8+16384 a b^2 \alpha  r_h^8\nonumber \\ &&+5391360 b^4 P \alpha  r_h^8+1096704 b^3  \alpha  r_h^8+30720 b P \alpha ^3 r_h^7+1536  \alpha ^3 r_h^7-61440 a b \alpha ^2 r_h^7+1276416 b^3 P \alpha ^2 r_h^7\nonumber \\ &&+248320 b^2  \alpha ^2 r_h^7+497664 b^6  r_h^7-62464 a b^3 \alpha  r_h^7+2571264 b^5 P \alpha  r_h^7+1465344 b^4  \alpha  r_h^7+192 P \alpha ^4 r_h^6\nonumber \\ &&-2688 a \alpha ^3 r_h^6+80640 b^2 P \alpha ^3 r_h^6+9984 b  \alpha ^3 r_h^6-98496 a b^2 \alpha ^2 r_h^6+1031616 b^4 P \alpha ^2 r_h^6+418560 b^3  \alpha ^2 r_h^6\nonumber \\ &&-95232 a b^4 \alpha  r_h^6+497664 b^6 P \alpha  r_h^6+964224 b^5  \alpha  r_h^6+1152 b P \alpha ^4 r_h^5-7168 a b \alpha ^3 r_h^5+103680 b^3 P \alpha ^3 r_h^5\nonumber \\ &&+25344 b^2  \alpha ^3 r_h^5-84352 a b^3 \alpha ^2 r_h^5+414720 b^5 P \alpha ^2 r_h^5+371520 b^4  \alpha ^2 r_h^5-36864 a b^5 \alpha  r_h^5+248832 b^6  \alpha  r_h^5\nonumber \\ &&+2592 b^2 P \alpha ^4 r_h^4-6704 a b^2 \alpha ^3 r_h^4+64800 b^4 P \alpha ^3 r_h^4+31104 b^3  \alpha ^3 r_h^4-37696 a b^4 \alpha ^2 r_h^4+62208 b^6 P \alpha ^2\nonumber \\ && r_h^4+156384 b^5  \alpha ^2 r_h^4+240 a b \alpha ^4 r_h^3+2592 b^3 P \alpha ^4 r_h^3-3200 a b^3 \alpha ^3 r_h^3+15552 b^5 P \alpha ^3 r_h^3 +24 a b^4 \alpha ^4\nonumber \\ && +18144 b^4  \alpha ^3 r_h^3-6144 a b^5 \alpha ^2 r_h^3+20736 b^6  \alpha ^2 r_h^3+364 a b^2 \alpha ^4 r_h^2+972 b^4 P \alpha ^4 r_h^2-720 a b^4 \alpha ^3 r_h^2\nonumber \\ &&+3888 b^5  \alpha ^3 r_h^2+172 a b^3 \alpha ^4 r_h\bigg]^{-1} \ .
\end{eqnarray}

\begin{figure}[]
 \begin{center}
 \subfigure[]{
 \includegraphics[height=4.7cm,width=7cm]{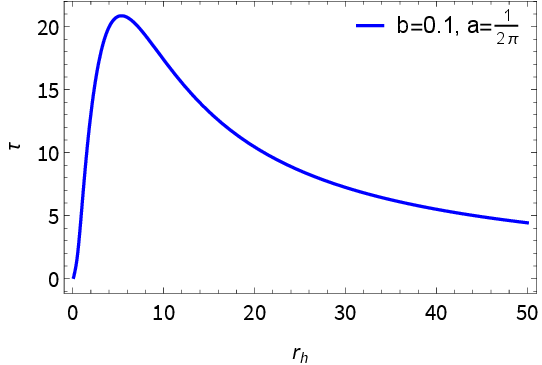}
 \label{300a}}
 \subfigure[]{
 \includegraphics[height=4.7cm,width=7cm]{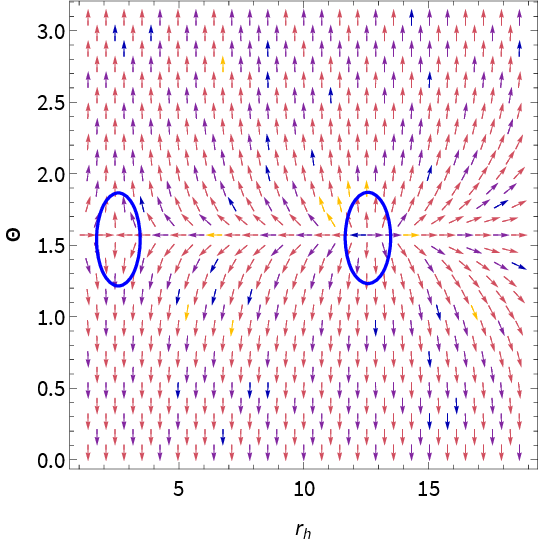}
 \label{300b}}\\
 \subfigure[]{
 \includegraphics[height=4.7cm,width=7cm]{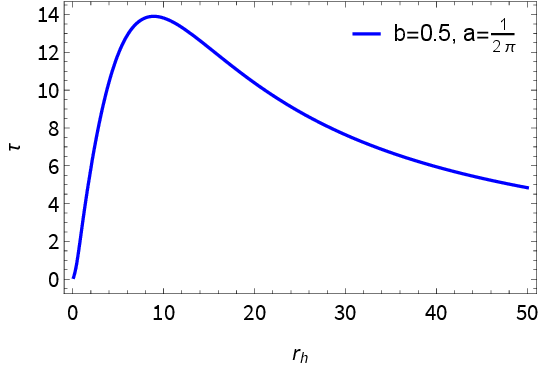}
 \label{300c}}
 \subfigure[]{
 \includegraphics[height=4.7cm,width=7cm]{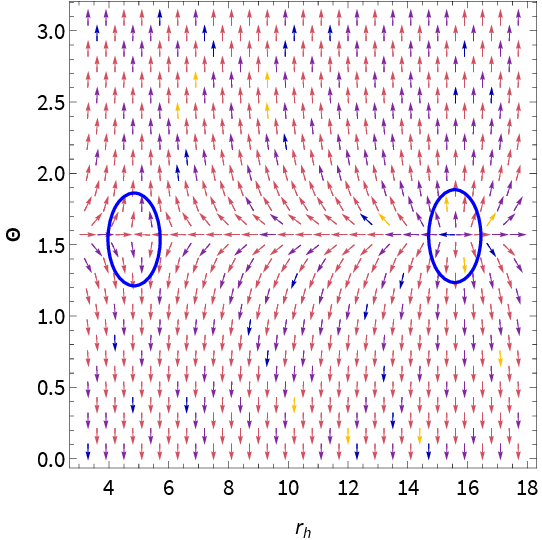}
 \label{300d}}\\
 \subfigure[]{
 \includegraphics[height=4.7cm,width=7cm]{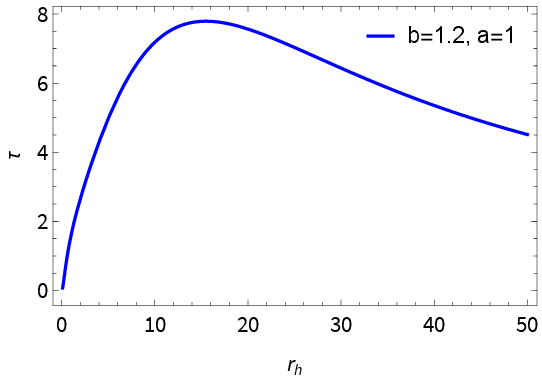}
 \label{300e}}
 \subfigure[]{
 \includegraphics[height=4.7cm,width=7cm]{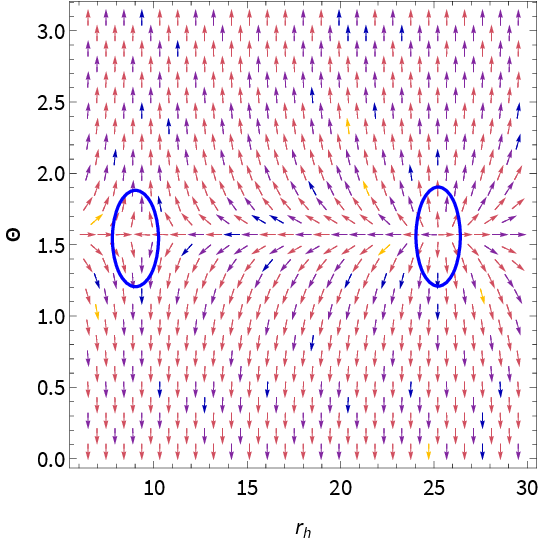}
 \label{300f}}
 \subfigure[]{
 \includegraphics[height=4.7cm,width=7cm]{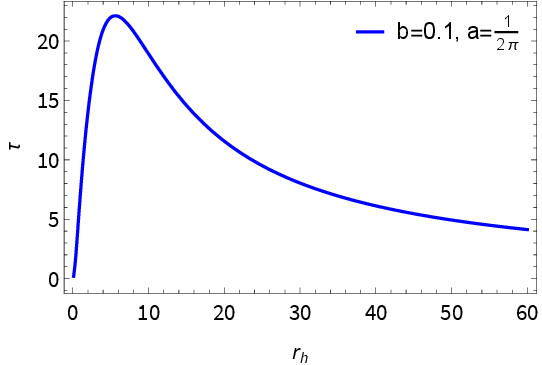}
 \label{300h}}
 \subfigure[]{
 \includegraphics[height=4.7cm,width=7cm]{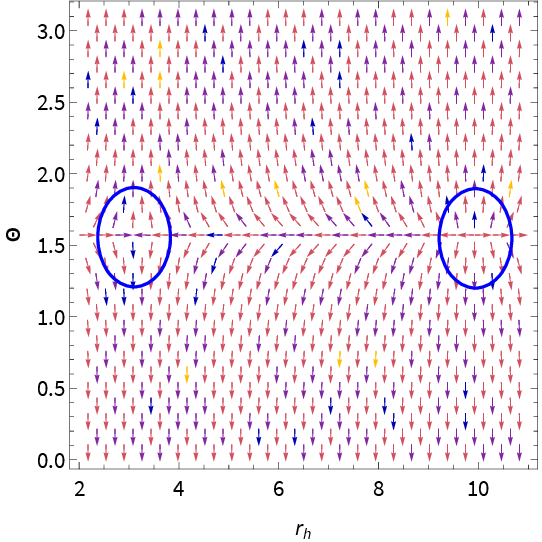}
 \label{300i}}
 \caption{\small{The $(\tau \text{ vs. } r_h)$ diagram, illustrating variations in free parameters for GUP corrected black hole, is presented in Figs. (\ref{300a}), (\ref{300c}), (\ref{300e}) and (\ref{300h}). Additionally, the normal vector field \( n \) in the \((r_h - \Theta)\) plane is depicted, with Zero Points (ZPs) located at specific coordinates \((r_h, \Theta)\). These ZPs correspond to parameter values $(\alpha=1, b = 0.1, a= 1/2\pi), (\alpha=1, b=0.5, a= 1/2\pi), (\alpha=1, b=1.2, a=1)$ and $(\alpha=0.5, b=0.1, a= 1/2\pi) $ along with fixed values of \( l= 1 \), as shown in Figs. (\ref{300b}), (\ref{300d}), (\ref{300f}) and (\ref{300i}), respectively.}}
 \label{m3}
\end{center}
 \end{figure}

Now, we examine the thermodynamic topology of the GUP-corrected black hole, focusing on the distribution of topological charges as illustrated in Fig.~\ref {m3}. The normalized field lines presented in the figure provide a visual representation of these characteristics. Fig.~(\ref{m3}) highlights two distinct zero points located at specific coordinates \((r_h, \Theta)\), corresponding to parameter values \( b = 0.1, 0.5, 1.2 \), \( \alpha = 0.5, 1 \), \( l = 1 \), and \( a = 1/2\pi, 1 \). These zero points, indicative of topological charges, are enclosed within blue contour loops, with their configuration shaped by the variations in free parameters. A crucial observation emerges from this analysis: when all free parameters—\( \alpha \), \( a \), and \( b \)—even are changed, we face two topological charges, \( (\omega = +1, -1) \) and the classification and topological charges even by changing the free parameters remain constat, resulting in a total topological charge of \( W = 0 \), as depicted in Fig.~\ref{m3}. The stability of the black hole is further examined through winding number calculations.

\newpage
\subsection{Thermodynamic Topology in EUP corrected vdW- Black holes}

\begin{figure}[]
 \begin{center}
 \subfigure[]{
 \includegraphics[height=3.4cm,width=3.4cm]{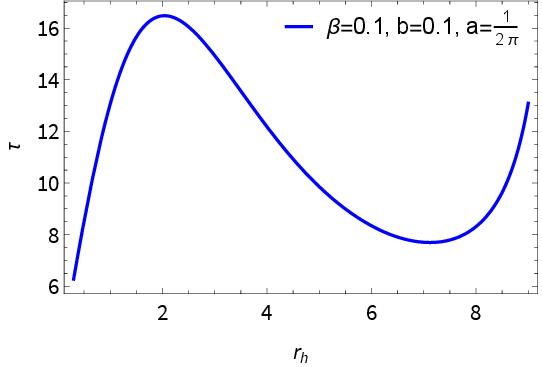}
 \label{500a}}
 \subfigure[]{
 \includegraphics[height=3.4cm,width=3.4cm]{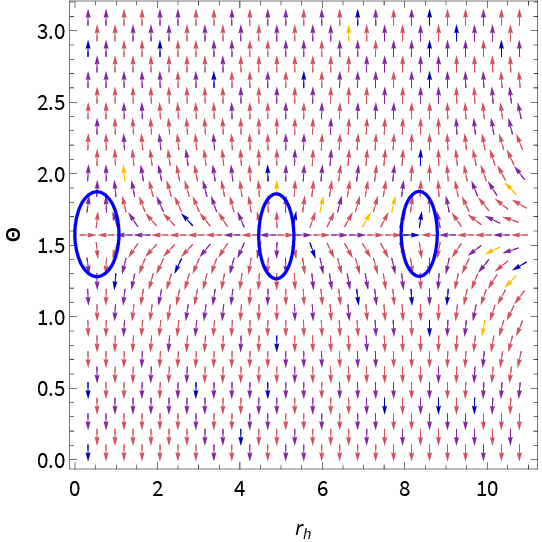}
 \label{500b}}
 \subfigure[]{
 \includegraphics[height=3.4cm,width=3.4cm]{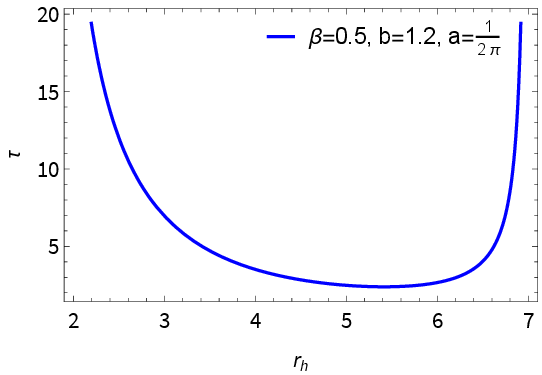}
 \label{500c}}
 \subfigure[]{
 \includegraphics[height=3.4cm,width=3.4cm]{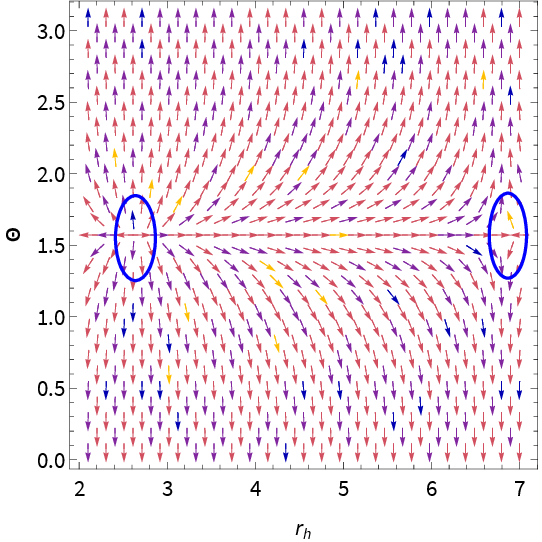}
 \label{500d}}
 \subfigure[]{
 \includegraphics[height=3.4cm,width=3.4cm]{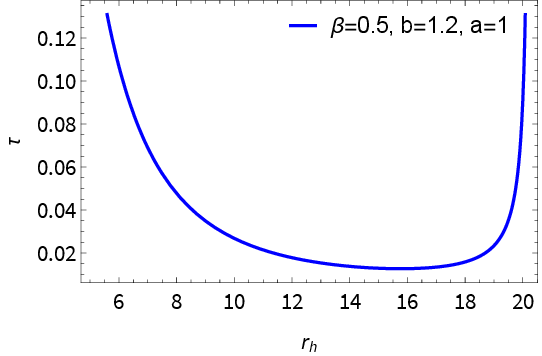}
 \label{500d2}}
 \subfigure[]{
 \includegraphics[height=3.4cm,width=3.4cm]{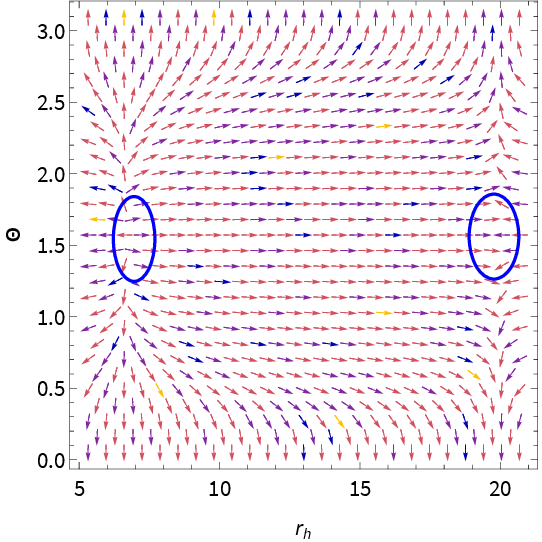}
 \label{500e}}\\
 \subfigure[]{
 \includegraphics[height=3.4cm,width=4.2cm]{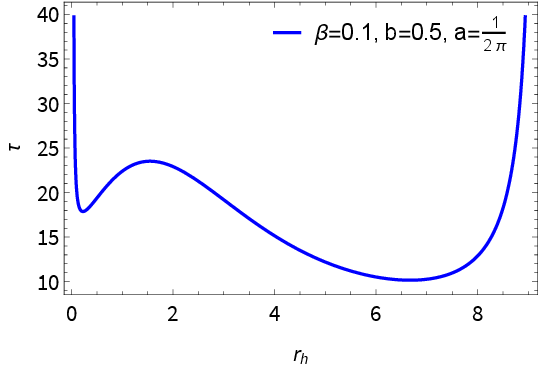}
 \label{500f}}
 \subfigure[]{
 \includegraphics[height=3.4cm,width=4.2cm]{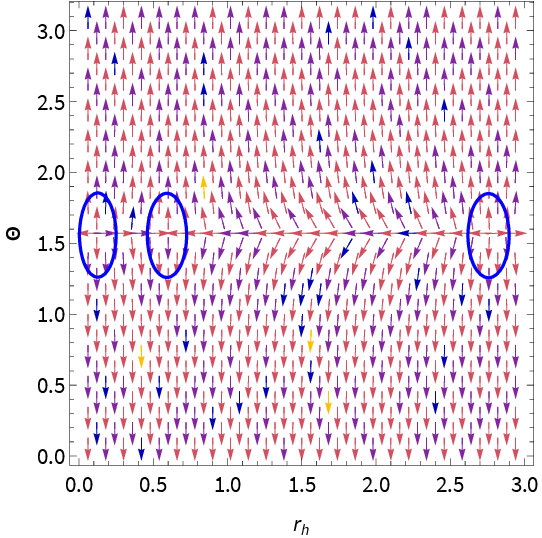}
 \label{500f2}}
 \subfigure[]{
 \includegraphics[height=3.4cm,width=4.2cm]{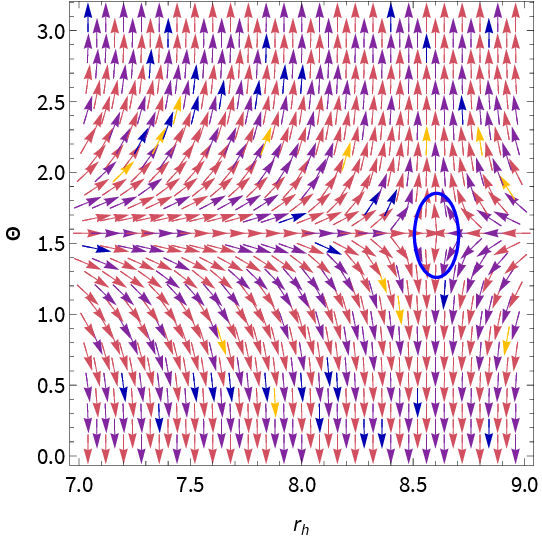}
 \label{500h}}\\
 \subfigure[]{
 \includegraphics[height=3.4cm,width=4.2cm]{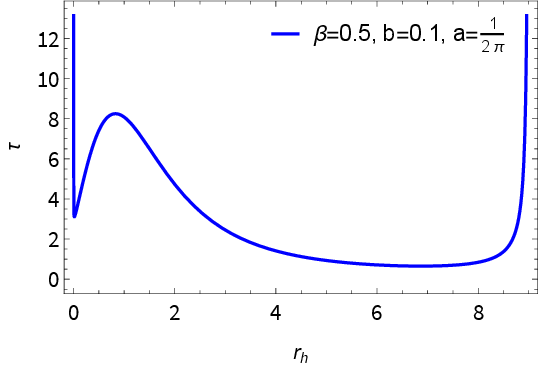}
 \label{500i}}
 \subfigure[]{
 \includegraphics[height=3.4cm,width=4.2cm]{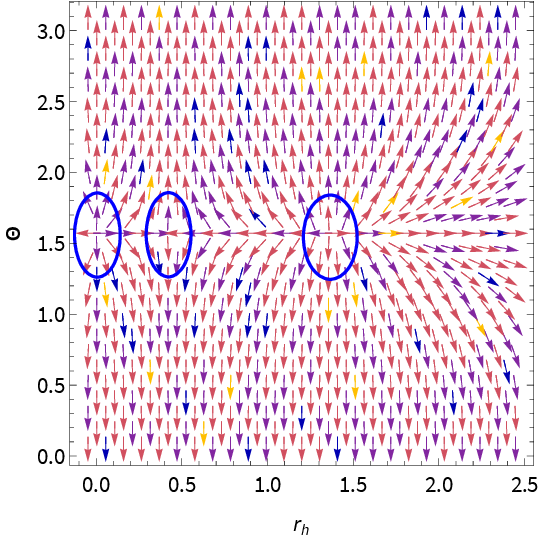}
 \label{500i2}}
 \subfigure[]{
 \includegraphics[height=3.4cm,width=4.2cm]{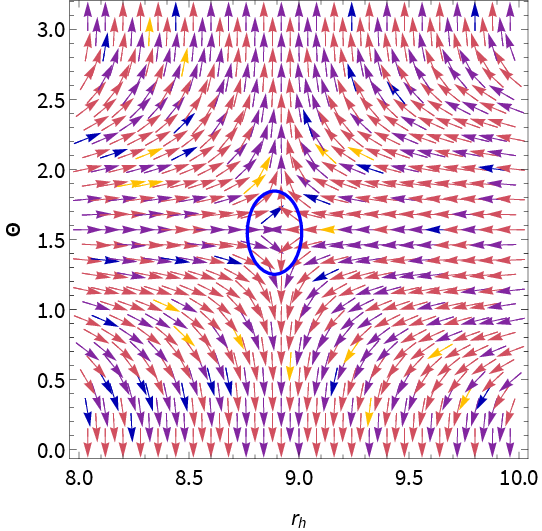}
 \label{500j}}\\
 \subfigure[]{
 \includegraphics[height=3.4cm,width=4.2cm]{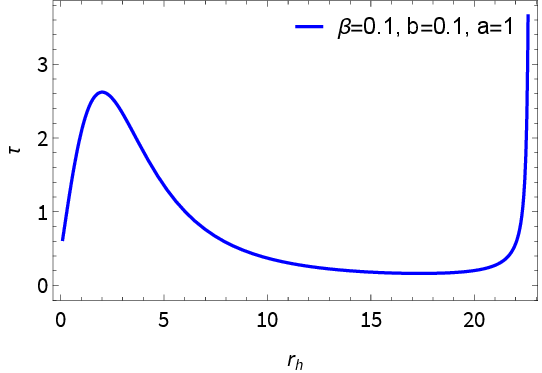}
 \label{500k}}
 \subfigure[]{
 \includegraphics[height=3.4cm,width=4.2cm]{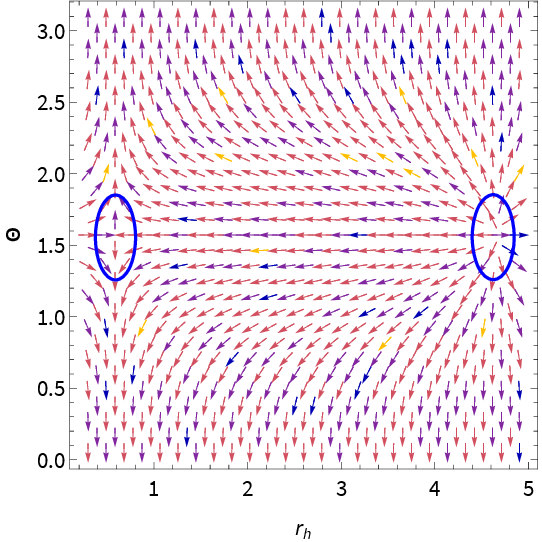}
 \label{500l}}
 \subfigure[]{
 \includegraphics[height=3.4cm,width=4.2cm]{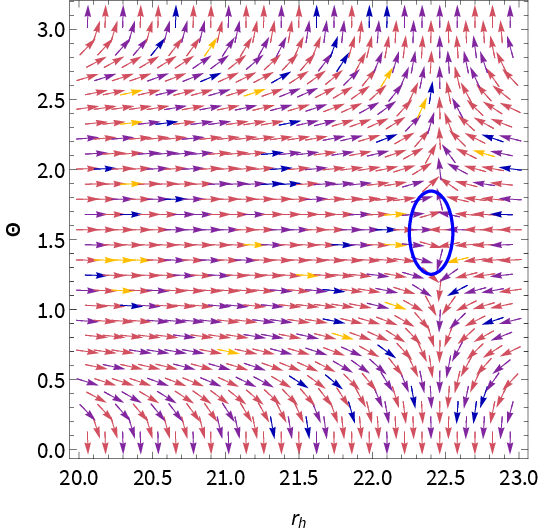}
 \label{500m}}
 \caption{\small{The $(\tau \text{ vs. } r_h)$ diagram, illustrating the influence of varying free parameters on EUP-corrected black holes, is presented in Figs. (\ref{500a}), (\ref{500c}), (\ref{500d2}), (\ref{500f}), (\ref{500i}), and (\ref{500k}). Furthermore, the normal vector field \( n \) is depicted in the \((r_h - \Theta)\) plane, where Zero Points (ZPs) appear at specific coordinates \((r_h, \Theta)\). These ZPs correspond to parameter values \( \beta = 0.1, 0.5 \), \( b = 0.1, 0.5, 1.2 \), and \( a = 1/2\pi, 1 \), providing a structured representation of the system’s topological behavior.}}
 \label{m5}
\end{center}
 \end{figure}

In this subsection, we examine the structure and classification of topological charges. Building upon Eq. (\ref{T1}), we establish the formulation of the generalized Helmholtz free energy for EUP-corrected black holes, highlighting their fundamental thermodynamic properties.
\begin{eqnarray}\label{E1}
\mathcal{F} &=&\frac{\pi}{48}  \Bigg[-\frac{243 a \beta  b^5 r_h}{3 b+2 r_h}+54 a \beta  b^2 r_h+\frac{9 a b^2 (45 b^2 \beta -8)}{3 b+2 r_h}+24 b r_h^2 (4 P-a \beta )-96 a b \log \left(\frac{r_h}{b}+\frac{3}{2}\right) \nonumber \\
&& \;\;\;\;\;\;\;\;\;\;\;\;\;\;\; +8 a \beta  r_h^3+48 a r_h-96 \beta  P r_h^4 (2 b+r_h)+64 P r_h^3-\frac{48 \log (\beta  r_h^2+1)}{\beta  \tau }\Bigg] \ .
\end{eqnarray}
We can calculate the $\phi^{r_h}$ and $\phi^\Theta$ with respect to Eq. (\ref{T2}) as follows,
\begin{eqnarray}\label{E2}
&&\phi^{r_h}=\frac{\pi  a (\beta  (72 b^3 r_h-48 b^2 r_h^2-27 (9 b^2+4) b^4+32 b r_h^3+32 r_h^4)+64 r_h (b+r_h))}{16 (3 b+2 r_h)^2} \nonumber \\
&& +2 \pi  r_h \left(\beta  (-P) r_h^2 (8 b+5 r_h)+2 P (b+r_h)-\frac{1}{\beta  r_h^2 \tau +\tau }\right) \;\;\;\;\;\;\;;\;\;\;\;\;\; \phi ^{\Theta }=-\frac{\cot (\Theta )}{\sin (\Theta )} \ .
\end{eqnarray}
We can calculate the $\tau$ of this mentioned black hole as follows,
\begin{equation}\label{E4}
\begin{split}
&\tau =-\bigg[32 r_h (3 b+2 r_h)^2\bigg]\times\bigg((\beta  r_h^2+1) (a \beta  (-72 b^3 r_h+48 b^2 r_h^2+27 (9 b^2+4) b^4-32 b r_h^3-32 r_h^4)\\&-64 a r_h (b+r_h)+32 P r_h (3 b+2 r_h)^2 (\beta  r_h^2 (8 b+5 r_h)-2 (b+r_h)))\bigg)^{-1}.
\end{split}
\end{equation}
This section examines the thermodynamic topology of EUP-corrected black holes, focusing on the distribution of topological charges as illustrated in Fig. (\ref{m5}). The normalized field lines presented in the figure provide a graphical depiction of these characteristics, offering valuable insights into their structural behavior.
Fig. (\ref{m5}) identifies several zero points at specific coordinates \((r_h, \Theta)\), corresponding to parameter values \( b = 0.1, 0.5, 1.2 \), \( \beta = 0.1, 0.5 \), and \( a = 1/2\pi, 1 \). These zero points, indicative of topological charges, are enclosed within blue contour loops, whose configurations are shaped by variations in free parameters.
It is evident that Figs. (\ref{500f2}) and (\ref{500h}), (\ref{500i2}) and (\ref{500j}), as well as (\ref{500l}) and (\ref{500m}), are grouped together. The separation of these Figs was necessary, as the spacing between the points did not accurately determine the zero points of the function.
In this model, we encounter two classifications of black holes based on different topological charges. For example, in Fig. (\ref{500b}), we observe three topological charges ($\omega=-1, +1, -1$) and a total topological charge of ($W=-1$) for the free parameters ($b=0.1, a=1/2\pi, \beta=0.1$). A similar classification, with identical topological charges, appears in Fig. (\ref{500l}+\ref{500m}), corresponding to the free parameters ($b=0.1, a=1, \beta=0.1$). 
In the remaining structures, the total topological charge is generally ($W=0$), and the differences between the figures primarily lie in the number of topological charges. Specifically, in Figs. (\ref{500d}) and (\ref{500e}), we observe two topological charges, resulting in a total topological charge of ($W=0$). In contrast, in Figs. (\ref{500f2}+\ref{500h}) and (\ref{500i2}+\ref{500j}), four topological charges are present, maintaining a total topological charge of ($W=0$).
Within this model, the role of the parameters ($b, a, \beta$) is clearly defined, distinguishing it from other models under study. Moreover, the free parameters for each diagram are explicitly specified. Consequently, this model presents two distinct classifications for the black holes under consideration.
\newpage
\subsection{Thermodynamic Topology in Rainbow-gravity corrected vdW- Black holes}

In this section, we analyze the classification of topological charges and their underlying structure. Utilizing Eq. (\ref{T1}) as a foundation, we derive the formulation of the generalized Helmholtz free energy for the Rainbow Gravity-corrected black hole, capturing its essential thermodynamic characteristics.
\begin{equation}\label{R1}
\begin{split}
&\mathcal{F}=\frac{\pi}{18}   \Bigg[\frac{a\gamma  (3 b+2 r_h) (45 b^2+30 b r-29 r_h)}{E_p^2}-\frac{27 a b^2}{3 b+2 r_h}+18a r_h+\frac{6 r_h^2 (6 b P \tau +4 P r_h \tau -3)}{\tau }\\&\;\;\;\;\;\;\;\;\-\frac{6 \gamma  P (2 b+3 r_h)}{E_p^2}\Bigg]-\frac{\gamma  (\log (r_h) (27 b-8 E_p^4 \pi  a \tau )+8 E_p^4 \pi  a \tau  \log (3 b+2 r_h))}{27 E_p^2 b \tau }-2 \pi  a b \log \left(\frac{r_h}{b}+\frac{3}{2}\right) \ ,
\end{split}
\end{equation}

\begin{figure}[]
 \begin{center}
 \subfigure[]{
 \includegraphics[height=3.8cm,width=3.8cm]{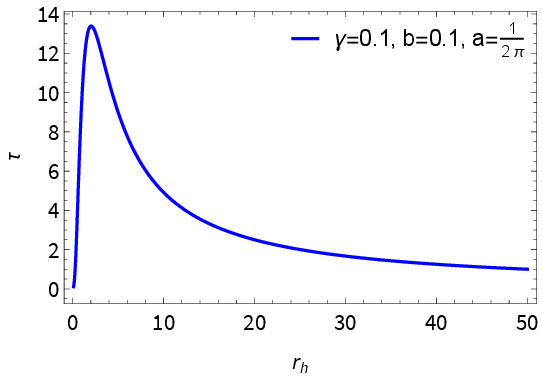}
 \label{400a}}
 \subfigure[]{
 \includegraphics[height=3.8cm,width=3.8cm]{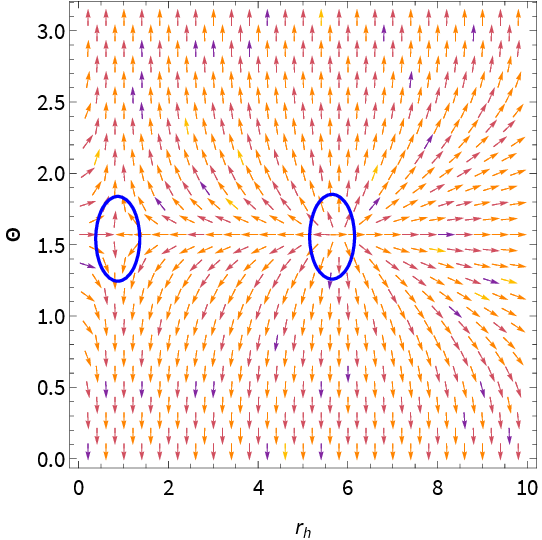}
 \label{400b}}
 \subfigure[]{
 \includegraphics[height=3.8cm,width=3.8cm]{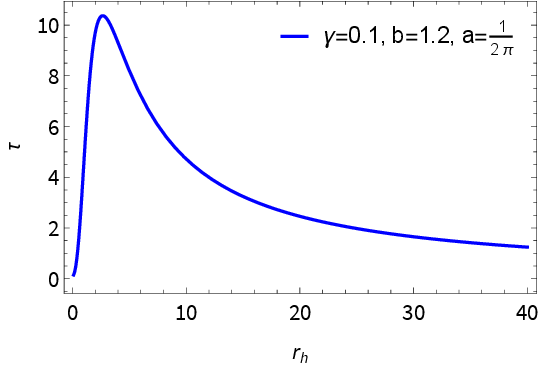}
 \label{400c}}
 \subfigure[]{
 \includegraphics[height=3.8cm,width=3.8cm]{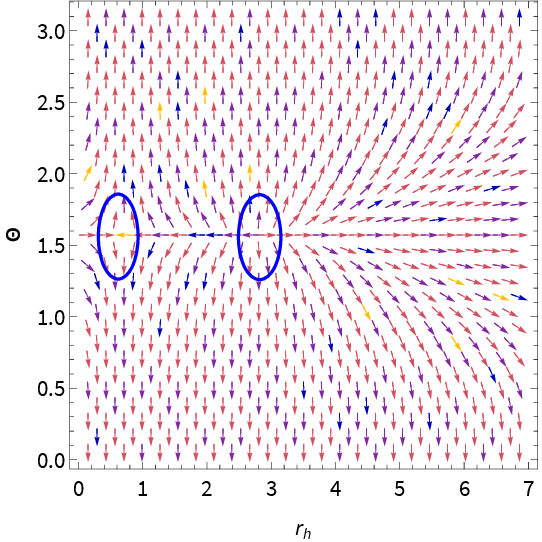}
 \label{400d}}
 \subfigure[]{
 \includegraphics[height=3.8cm,width=3.8cm]{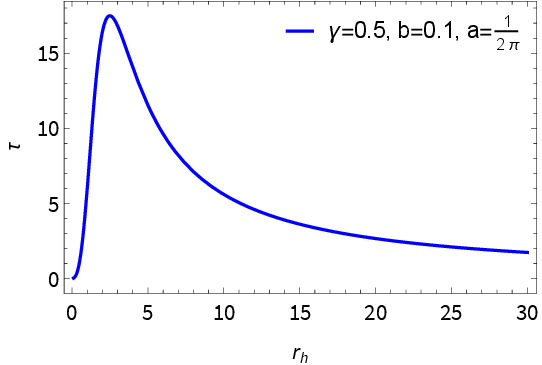}
 \label{400e}}
 \subfigure[]{
 \includegraphics[height=3.8cm,width=3.8cm]{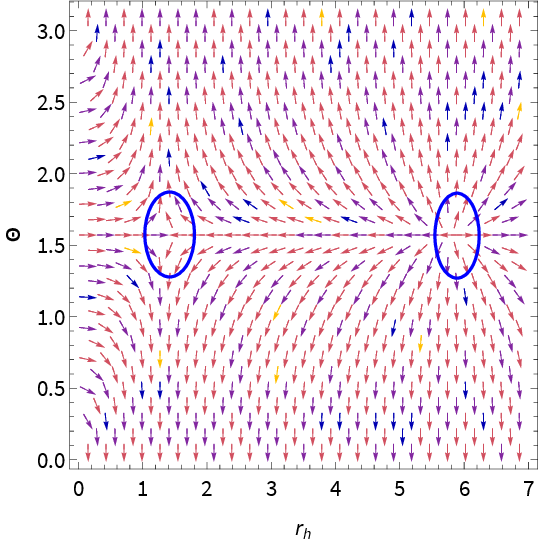}
 \label{400f}}
 \subfigure[]{
 \includegraphics[height=3.8cm,width=3.8cm]{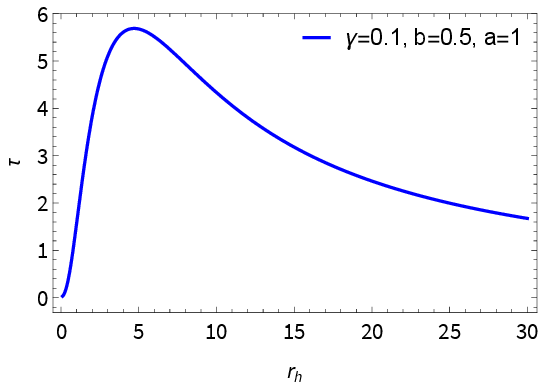}
 \label{400h}}
 \subfigure[]{
 \includegraphics[height=3.8cm,width=3.8cm]{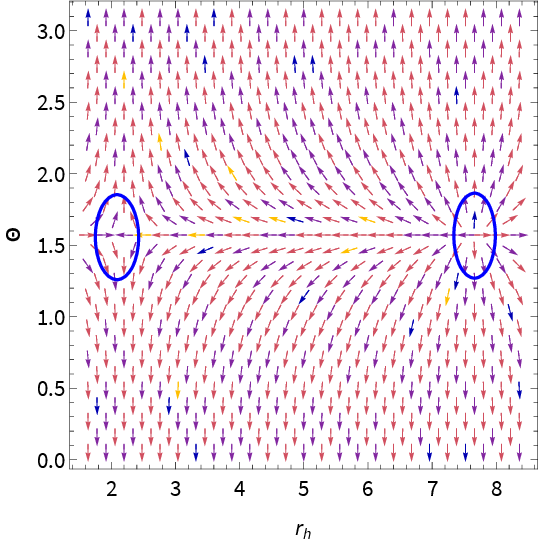}
 \label{400i}}
 \subfigure[]{
 \includegraphics[height=3.8cm,width=3.8cm]{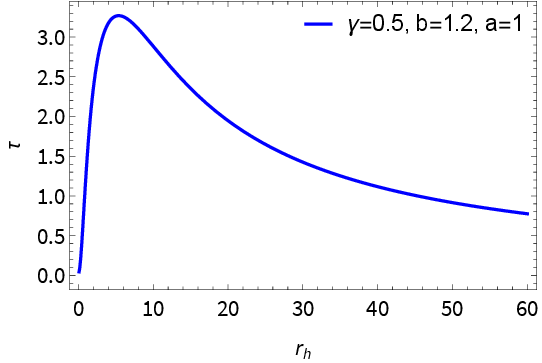}
 \label{400j}}
 \subfigure[]{
 \includegraphics[height=3.8cm,width=3.8cm]{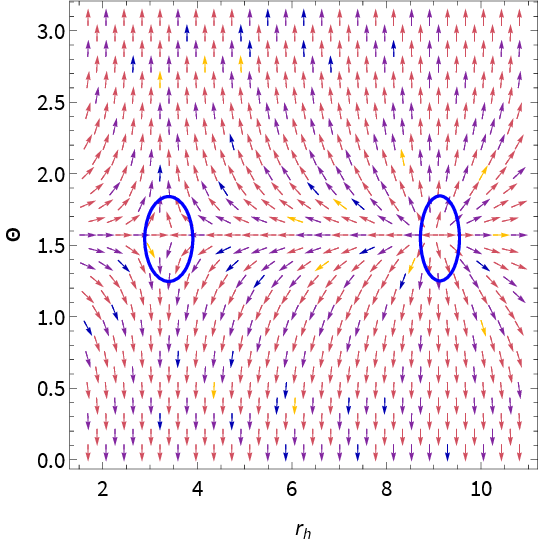}
 \label{400k}}
 \caption{\small{The $(\tau \text{ vs. } r_h)$ diagram, illustrating variations in free parameters for Rainbow Gravity-corrected black hole, is presented in Figs. (\ref{400a}), (\ref{400c}), (\ref{400e}), (\ref{400h}) and (\ref{400j}). Additionally, the normal vector field \( n \) in the \((r_h - \Theta)\) plane is depicted, with Zero Points (ZPs) located at specific coordinates \((r_h, \Theta)\). These ZPs correspond to parameter values $(\gamma=0.1, 0.5, b = 0.1, 0.5, 1.2, a= 1/2\pi, a)$.}}
 \label{m4}
\end{center}
 \end{figure}

With respect to Eq. (\ref{T2}), $\phi^{r_h}$ and $\phi^\Theta$ can be determined as,
\begin{equation}\label{R3}
\begin{split}
\phi^{r_h}=\frac{\mathcal{X}+\mathcal{Y}}{18 r} \;\;\;\;\;\;\;\;\;\;;\;\;\;\;\;\;\;\;\; \phi ^{\Theta }=-\frac{\cot (\Theta )}{\sin (\Theta )}
\end{split}
\end{equation}
where
\begin{equation}\label{R2}
\begin{split}
&\mathcal{X}=\frac{\pi  a}{(3 b+2 r_h)^2} \left(72 r_h^2 (b+r_h)+\frac{\gamma  (3 b+2 r_h) (r_h (3 b+2 r_h) (3 b (60 b+40 r_h-29)-116 r_h)+16 E_p^4)}{E_p^2}\right) \ , \\
&\mathcal{Y}=\frac{18 }{\tau }\left(2 \pi  r_h^2 (2 P \tau  (b+r_h)-1)-\frac{\gamma +\pi  \gamma  P r_h \tau }{E_p^2} \right) \ .
\end{split}
\end{equation}
Also, by solving them, we have
\begin{eqnarray}\label{R4}
    &&\tau =\frac{\left(18 (3 b+2 r_h)^2 (\gamma +2_h E_p^2 \pi  r_h^2)\right)}{\pi}\bigg[1620 a b^4 \gamma  r_h+3240 a b^3 \gamma  r_h^2-783 a b^3 \gamma  r_h+2160 a b^2 \gamma  r_h^3+48 E_p^4 a b \gamma \nonumber \\
    &&-2088 a b^2 \gamma  r_h^2 +480 a b \gamma  r_h^4-1740 a b \gamma  r_h^3+72 E_p^2 a b r_h^2-464 a \gamma  r_h^4+72 E_p^2 a r_h^3+32 E_p^4 a \gamma  r_h -162 b^2 \gamma  P r_h \nonumber \\
    &&+648 E_p^2 b^3 P r_h^2+1512 E_p^2 b^2 P r_h^3+1152 E_p^2 b P r_h^4-216 b \gamma  P r_h^2+288 E_p^2 P r_h^5-72 \gamma  P r_h^3 \bigg]^{-1} \ .
\end{eqnarray}
The thermodynamic topology of the Rainbow Gravity-corrected black hole, emphasizing the distribution of topological charges as illustrated in Fig. (\ref{m4}). The normalized field lines depicted in the figure provide a graphical representation of these characteristics, offering insight into their structural behavior. Fig. (\ref{m4}) reveals two distinct zero points at specific coordinates \((r_h, \Theta)\), corresponding to parameter values \( b = 0.1, 0.5, 1.2 \), \( \gamma = 0.1, 0.5 \), and \( a = 1/2\pi, 1 \). These zero points, associated with topological charges, are encased within blue contour loops, whose configurations are influenced by variations in free parameters.
A key finding from this analysis is that, even when the free parameters—\(\gamma\), \( a \), and \( b \)—are varied, the system consistently exhibits two topological charges \( (\omega = +1, -1) \). The classification and presence of these charges remain unchanged despite modifications in the free parameters, leading to a total topological charge of \( W = 0 \), as depicted in Fig. (\ref{m4}). Furthermore, the stability of the black hole is assessed through winding number calculations, providing additional insights into its topological and thermodynamic properties.\\

In the study of black hole thermodynamics, one powerful approach involves visualizing the free energy as a scalar field distributed over a two-dimensional parameter space. This space is typically defined by the black hole’s horizon radius and a thermodynamic control variable, such as temperature, pressure, or electric potential. By examining how the free energy changes across this space, one can construct a corresponding vector field that captures the direction and rate of change—essentially, the thermodynamic “flow” of the system. Within this framework, the extremum points of the free energy—locations where the system is in thermodynamic equilibrium—emerge as zero points of the vector field. These are the points where the gradient of the free energy vanishes, indicating no net thermodynamic force acting on the system. The nature of these points—whether they represent stable, unstable, or transitional states—can be inferred by analyzing the behavior of the vector field in their vicinity. Around each zero point, the vector field lines typically exhibit a swirling or rotational pattern. This rotational behavior is not random; it encodes meaningful information about the local stability of the configuration. To quantify this, one uses the concept of the winding number, a topological invariant that counts how many times the vector field rotates as one encircles the zero point along a closed path. The direction and number of these rotations determine the winding number’s value. A winding number of +1 generally corresponds to a stable equilibrium, such as a local minimum in the free energy landscape. A winding number of –1 indicates an unstable equilibrium, often associated with a local maximum. A winding number of 0 suggests a neutral or saddle-like configuration, where the system may be sensitive to perturbations but lacks a clear stability signature. Each winding number serves as a local topological charge, characterizing the thermodynamic behavior at a specific point in the parameter space. When all such local charges are summed across the entire domain, the result is the total topological charge of the system. This global quantity provides a robust classification of the black hole’s overall thermodynamic phase structure, independent of the specific details of the underlying geometry or matter content. Empirical studies of well-known black hole solutions have revealed a consistent pattern in these topological charges: The Schwarzschild black hole, which lacks charge or rotation, exhibits a total topological charge of –1, reflecting its inherently unstable thermodynamic nature. The Reissner–Nordström black hole, which includes electric charge, has a total charge of 0, indicating a balance between stable and unstable phases. The AdS–Reissner–Nordström black hole, embedded in anti-de Sitter spacetime, shows a total charge of +1, signifying the presence of a globally stable phase.
These findings underscore the power of topological analysis in black hole thermodynamics. Rather than focusing on the precise form of the metric or the specific dynamics of the matter fields, this approach captures the qualitative structure of the thermodynamic landscape. The total topological charge acts as a global diagnostic, while the individual winding numbers offer localized insight into the stability and phase behavior of the system.
\subsection{Photon Sphere in vdW black holes}


To discuss the photon sphere for these black holes, we start with the form of metric as in \eqref{Metric Ansatz} and discussed in detail in \cite{45m,46m,47m,48m,49m}. Our aim is to analyze the null geodesics and determine the location of the photon sphere, which requires an effective potential. Due to the inherent \( Z_2 \) symmetry, the problem can be simplified by restricting the analysis to the equatorial plane (\( \theta = \pi/2 \)), without loss of generality. In general, after some calculations, the geodesic equation takes the form,
\begin{equation}\label{ph2}
\dot{r}^2 + V_{\text{eff}} = 0.
\end{equation}
The effective potential, \( V_{\text{eff}} \), is expressed as,
\begin{equation}\label{ph3}
V_{\text{eff}} = g(r) \left( \frac{L^2}{r^2} - \frac{E_p^2}{f(r)} \right) \ , 
\end{equation}
where \( E \) and \( L \) denote the photon's energy and angular momentum, respectively, corresponding to the Killing vector fields \( \partial_t \) and \( \partial_\phi \). Due to spherical symmetry, the photon sphere is located at a radial coordinate \( r_{\text{ps}} \), where the following conditions hold,
\begin{equation}\label{ph4}
V_{\text{eff}} = 0 \, , \quad \frac{dV_{\text{eff}}}{dr} = 0 \ .
\end{equation}
This leads to the condition $\left( r^{-2}f(r) \right)'|_{r = r_{\text{ps}}} = 0,$ where the prime denotes differentiation with respect to \( r \). The stability of the photon sphere is determined by the second derivative,
\( \frac{d^2 V_{\text{eff}}}{dr^2} < 0 \) for an unstable photon sphere. \( \frac{d^2 V_{\text{eff}}}{dr^2} > 0 \) for a stable photon sphere. At the black hole horizon \( r_h \), where \( f(r_h) = 0 \), the first term vanishes, leaving the second term generally nonzero. This implies that, in general, \( r_{\text{ps}} \neq r_h \). However, in the case of an extremal black hole—where the two horizons merge—the conditions \( f(r_h) = 0 \) and \( f'(r_h) = 0 \) hold, leading to the coincidence of the photon sphere with the extremal horizon. Instead of the conventional approach, this study examines the photon sphere behavior through the topological method. In this framework, each photon sphere is characterized by its topological charge. Since this methodology has been extensively detailed and applied in various models, we focus on its essential relationships rather than repeating previous derivations. To explore the topology of the photon sphere, we introduce the regular potential function,
\begin{equation}\label{ph6}
H(r, \theta) = r - \frac{g_{tt}}{g_{\phi\phi}} = \frac{\sqrt{f(r)}}{r \sin \theta} \ .
\end{equation}
The photon sphere radius is obtained as the root of \( \frac{dH}{dr} = 0 \). Using the components of the vector field \( \phi = (\phi^{r_h}, \phi^\theta) \), we define,
\begin{equation}\label{ph7}
\phi^{r_h} = \frac{1}{\sqrt{g_{rr}}} \frac{dH}{dr} = \sqrt{f(r)} \frac{dH}{dr}, \quad \phi^\theta = \frac{1}{\sqrt{g_{\theta\theta}}} \frac{dH}{d\theta} = \frac{1}{r} \frac{dH}{d\theta}.
\end{equation}
Rewriting the vector field representation, we obtain, $\phi = \|\phi\| e^{i\Theta}, \quad \|\phi\| = \sqrt{\phi^a \phi_a}.$ Thus, the vector field can alternatively be expressed as $\phi = \phi^{r_h} + i \phi^\theta. $ The normalized vectors are defined as: $n^a = \frac{\phi^a}{\|\phi\|}, \quad \text{where} \quad (\phi^1 = \phi^{r_h},\, \phi^2 = \phi^\theta). $ According to Eqs. (\ref{ph6}) and  (\ref{ph7}) we have,
\begin{equation}\label{ph9}
    \begin{split}
        \phi^{r_h}=\frac{\mathcal{A}}{\mathcal{B}} \;\;\;\;\;;\;\;\;\;\; \phi^{\theta }=-\frac{\cot (\theta ) \csc (\theta ) \sqrt{-\frac{3 \pi  a b^2}{r_h (3 b+2 r_h)}-\frac{4 \pi  a b \log (\frac{r_h}{b}+\frac{3}{2})}{r_h}+2 \pi  a+\frac{r_h^2 (\frac{3 b}{2 r_h}+1)}{l^2}-\frac{2 M}{r_h}}}{r_h}
    \end{split}
\end{equation}
where
\begin{equation}\label{ph8}
\begin{split}
&\mathcal{A}=\csc (\theta ) (27 b^3 (2 \pi  a l^2-r_h^2)-36 b^2 (l^2 (2 \pi  a r_h-3 M)+r_h^3)-4 b r_h (4 l^2 (8 \pi  a r_h-9 M)+3 r_h^3)\\& \;\;\;\;\;\; \;\;\;\;\;\;\;\;\;\;\;\;\;\; \;\;\;\;\;\;\;\; +24 \pi  a b l^2 (3 b+2 r_h)^2 \log (\frac{r_h}{b}+\frac{3}{2})+16 l^2 r_h^2 (3 M-2 \pi  a r_h))\\
&\mathcal{B}=2 \sqrt{2} l^2 r_h^3 (3 b+2 r_h)^2\times\bigg(r_h^2 (8 \pi  a l^2+9 b^2)+4 l^2 r_h (3 \pi  a b-2 M)-6 b l^2 (\pi  a b+2 M)\\& \;\;\;\;\;\; \;\;\;\;\;\;\;\;\;\;\;\;\;\; \;\;\;\;\;\;\;\; -8 \pi  a b l^2 (3 b+2 r_h) \log (\frac{r_h}{b}+\frac{3}{2})+12 b r_h^3+4 r_h^4\bigg/l^2 r_h (3 b+2 r_h)\bigg) \ .
\end{split}
\end{equation}

From this, we examine the behavior of photon spheres (PS) within the framework of thermodynamic topology. A fundamental topological property states that a zero point of the vector field \( \vec{\phi} \) enclosed within a closed loop contributes a topological charge equal to the winding number. Each photon sphere is assigned a charge of either \( +1 \) or \( -1 \), depending on its winding direction. The total topological charge, determined by the chosen loop enclosing one or multiple zero points, can take values of \( -1 \), \( 0 \), or \( +1 \). In classical black hole configurations where \( M > Q \), the expected photon sphere structure leads to a total topological charge of \( -1 \).

\begin{figure}[]
 \begin{center}
 \subfigure[]{
 \includegraphics[height=5cm,width=5cm]{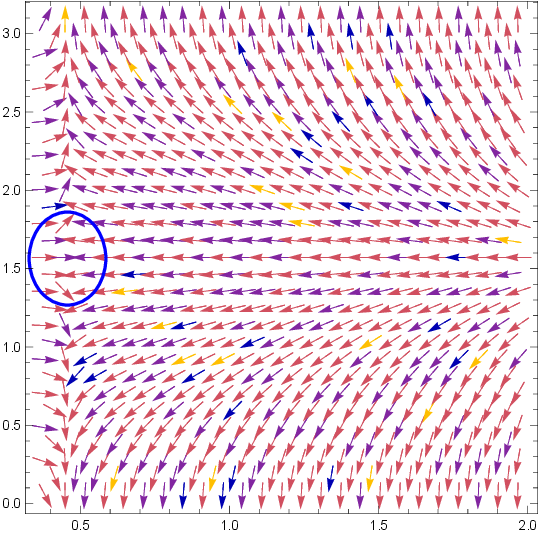}
 \label{200a}}
 \subfigure[]{
 \includegraphics[height=5cm,width=5cm]{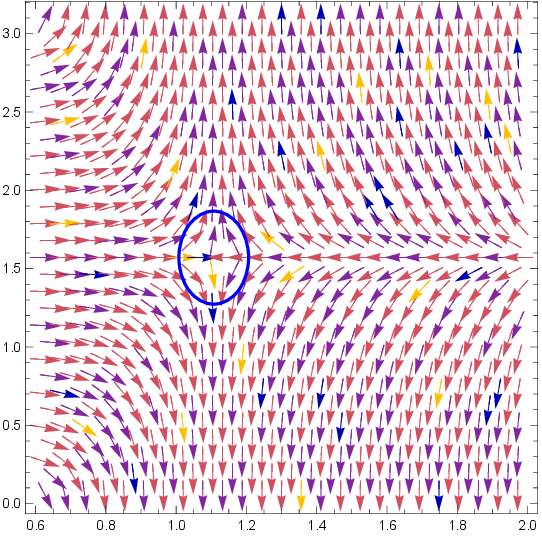}
 \label{200b}}
 \subfigure[]{
 \includegraphics[height=5cm,width=5cm]{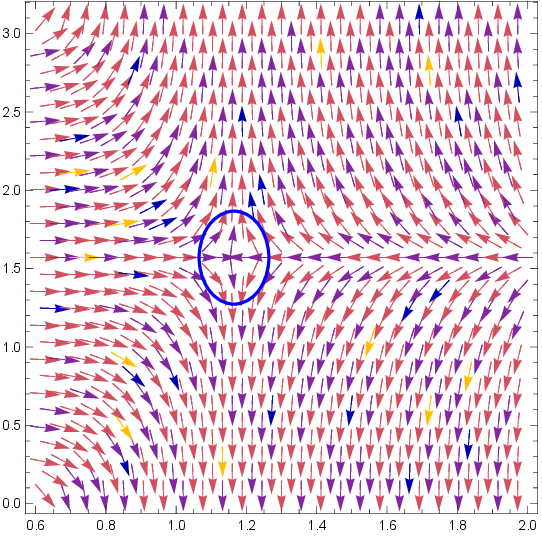}
 \label{200c}}
 \caption{\small{The plotted photon spheres (PSs) of van der Waals (vdW) black holes are presented in Fig. (\ref{200a}) for \( b = 0.1 \), Fig. (\ref{200b}) for \( b = 0.5 \), and Fig. (\ref{200c}) for \( b = 1.2 \). These visualizations correspond to the parameter values \( l = 1 \), \( a = 1/2\pi \), \( c = 8\pi/3 \), and \( M = 0.1 \)}}
 \label{m2}
\end{center}
 \end{figure}

To explore how this structure evolves under varying parameters, we analyze the photon sphere behavior for different values of the free parameters: \( l = 1 \), \( b = 0.1, 0.5, 1.2 \), \( a = 1/2\pi \), and \( c = 8\pi/3 \). The findings, illustrated in Fig. (\ref{m2}), indicate that across these parameter values, the system consistently exhibits a total photon sphere charge of \( -1 \). Investigating photon spheres provides deeper insights into the geometric and physical properties of the black holes considered. By examining the structure and effective geometry of photon orbits, we assess the influence of free parameters and \( b \) on photon sphere topology.

Our results reveal that unstable photon spheres exhibit significant structural features, offering deeper perspectives on black hole configurations. Moreover, the connection between black hole topological charges and photon sphere topology introduces new opportunities for studying gravitational lensing and shadow formation, with critical implications for astrophysical observations. Through comprehensive illustrations and mathematical derivations, we present a detailed analysis of how entropy parameter variations affect black hole thermodynamic topology. The results consistently show that positive winding numbers correspond to stable black hole configurations, while negative winding numbers indicate instability. These insights not only enhance our theoretical understanding but also provide a topological framework for interpreting black hole stability.

\section{Conclusion}

Extended black hole thermodynamics treats the cosmological constant as a thermodynamic pressure, thereby enabling the formulation of a consistent equation of state for black holes. This approach permits a direct analogy between black hole thermodynamic behavior and that of classical fluids, such as those described by the VdW equation. Over recent years, a variety of asymptotically AdS black hole solutions have been developed whose thermodynamic profiles exhibit striking correspondence with the VdW fluid, capturing critical phenomena and phase transitions. Additionally, quantum gravity corrections—particularly those arising from the generalized uncertainty principle (GUP), which introduces a fundamental minimal length—have been applied to VdW black hole models. These quantum modifications have been shown to significantly affect the thermodynamic structure, highlighting the important role of quantum deformations in black hole physics. The GUP corrections impose a physically meaningful domain for the event horizon radius, resolving issues such as negative horizon values encountered in the semi-classical analysis.  The Extended Uncertainty Principle (EUP) incorporates a fundamental lower bound on momentum uncertainty. The EUP framework introduces a natural upper limit on the black hole’s event horizon radius - a feature absent in conventional analyses. The entropy is modified under EUP, exhibiting a suppressed growth rate compared to the standard case. Stability analysis shows that the VdW black hole may transition between stable and unstable phases depending on the interplay of thermodynamic parameters. Rainbow Gravity modifies the spacetime metric itself by introducing energy-dependent functions, leading to modified dispersion relations. This effectively means that the geometry of spacetime seen by a particle depends on its energy, and thus, quantum effects alter gravitational backgrounds dynamically. The thermodynamic structure remains qualitatively consistent; the Rainbow Gravity framework introduces a lower bound on the event horizon radius, defining the threshold for physical and stable black hole configurations. Within the valid domain, the temperature increases monotonically, though with a suppressed magnitude compared to the classical case. The corrected entropy gains an additional positive term, reflecting a deviation from the standard area law. The significance of rainbow gravity as a viable framework for incorporating quantum gravity corrections into black hole thermodynamics, consistent with other approaches such as GUP and EUP.

In the case of Van der Waals (VdW) black holes, the entropy is the Bekenstein-Hawking entropy. This entropy satisfies the universality relation, a well-known result that applies to various black hole systems. The universality relation connects the entropy, mass, and thermodynamic variables, establishing a fundamental relationship between them across different types of black holes. However, when quantum gravitational effects, such as the GUP, the EUP, and Rainbow Gravity, are incorporated into the model, these corrections modify the thermodynamic properties of the black hole, including its entropy. To quantify the impact of these effects on the black hole entropy, we consider their contributions as small perturbations and treat the coupling constants associated with GUP, EUP, and Rainbow Gravity as small parameters. We perform a perturbative expansion of the event horizon radius, considering corrections up to first order in the coupling constants of these quantum effects. This allows us to derive modified expressions for the thermodynamic quantities. After incorporating these modifications, we verified the generalized universality relation. This relation extends the classical universality relation by accounting for quantum corrections, ensuring that the modified entropy still maintains a universal connection with other thermodynamic quantities, even in the presence of these quantum gravitational effects. Our results indicate that the generalized relation holds true even after introducing the corrections from GUP, EUP, and RaGr, thus confirming the robustness of the generalized universality relation in quantum-corrected black hole thermodynamics.

We also provided a comprehensive examination of the thermodynamic topology of various black hole models, highlighting the interplay between topological charge distributions and stability properties. The analysis spans multiple frameworks, including van der Waals (vdW) black holes, Generalized Uncertainty Principle (GUP)-corrected black holes, Extended Uncertainty Principle (EUP)-corrected black holes, and Rainbow Gravity-corrected black holes.  For vdW black holes, the findings illustrate how parameter variations influence topological charge distributions, leading to shifts in classification and stability transitions. The presence of multiple topological charges and their arrangement within contour loops demonstrate structural modifications dictated by changes in parameters. The study further extends to GUP-corrected black holes, revealing a persistent classification that remains unchanged despite modifications in free parameters. This robustness underscores the stability of topological charge distributions, reinforcing fundamental thermodynamic properties. In EUP-corrected vdW black holes, two distinct classifications emerge based on topological charge arrangements. The presence of different charge distributions highlights the influence of specific parameters in shaping black hole characteristics, distinguishing this model from other thermodynamic configurations. Rainbow Gravity-corrected black holes exhibit consistent topological charge distributions, demonstrating structural stability even with variations in free parameters. The classification remains intact, indicating a resilient framework in thermodynamic topology. Across all models, winding number evaluations serve as a critical tool in assessing black hole stability and classification transitions. The findings contribute to a deeper understanding of thermodynamic topology, offering insights into the role of parameter-driven transformations in black hole physics. This study provides a foundational perspective for future investigations into topological properties and their implications in theoretical physics.

For the Van der Waals black hole, we have identified the presence of unstable photon spheres, which serve as a feature of black hole structure. These photon spheres, governed by the intricate dynamics of null geodesics, provide crucial insights into the optical properties of the gravitational field surrounding the black hole. However, when examining the other three models in their normal form, we encounter significant analytical challenges in explicitly determining the photon spheres. Due to the complexity of the underlying equations, deriving closed-form expressions for the photon spheres becomes exceedingly difficult. In such cases, we resort to expanded relations to approximate their behavior. Nevertheless, upon doing so, it becomes evident that the roots of the function \(\phi^{r_h}\) lie within the event horizon of the black hole. Since these photon spheres are located within the horizon, they are effectively inaccessible to external observers and thus remain unobservable in a practical sense. Given these constraints, we opt to omit the explicit graphical representation of photon spheres for these models at this stage, as their physical relevance from an observational standpoint is limited.

This study presents a comprehensive exploration of how quantum gravitational effects—specifically those introduced by the Generalized Uncertainty Principle (GUP), the Extended Uncertainty Principle (EUP), and Rainbow Gravity—reshape our understanding of black hole thermodynamics. By incorporating these corrections, the work delves into how Planck-scale physics alters classical thermodynamic quantities such as entropy, temperature, and critical behavior. These modifications are not merely technical adjustments; they offer a more refined and physically meaningful description of black hole microstructure, particularly in regimes where classical theories become inadequate—such as near extremality or during the final stages of black hole evaporation. One of the central contributions of this study lies in its examination of universality relations associated with extremality bounds. Traditionally, extremal black holes are defined by conditions like zero temperature or the merging of inner and outer horizons. However, when quantum corrections are introduced, these extremal conditions may no longer hold in their classical form. Instead, they may shift or acquire new structural features, suggesting that the very definition of a terminal or stable black hole state must be reconsidered. This opens the door to novel possibilities, such as the existence of stable remnants or modified evaporation endpoints—scenarios that are of particular interest in quantum gravity and information loss debates. Complementing this thermodynamic analysis is the study’s use of topological tools, particularly the concepts of topological charge and winding number. These tools provide a geometric and coordinate-independent framework for classifying black hole phases. Rather than relying solely on thermodynamic potentials or response functions, the study maps the behavior of the free energy landscape as a vector field over a two-dimensional parameter space. The rotational patterns of this field around equilibrium points allow for the assignment of winding numbers, which serve as local indicators of stability. When summed across the entire space, these yield a total topological charge—a global signature of the system’s phase structure. This topological classification scheme offers a universal language for comparing different black hole solutions, including those modified by quantum corrections. It enables researchers to distinguish between stable, unstable, and transitional configurations in a way that is robust under smooth deformations of the system. Moreover, by analyzing the corrected free energy landscape, the study introduces new tools for identifying and characterizing phase transitions. These include not only the familiar first- and second-order transitions but also more subtle phenomena like zeroth-order transitions, which are often overlooked in classical treatments. The inclusion of GUP, EUP, and Rainbow Gravity in this analysis is particularly impactful. These frameworks introduce corrections that can shift critical points, alter coexistence curves, and even generate entirely new thermodynamic phases. Such features may remain hidden in purely classical models, underscoring the importance of quantum corrections in capturing the full richness of black hole behavior. What sets this work apart is its integrative approach. While previous studies have examined Van der Waals analogies or quantum corrections in isolation, this study brings them together within a unified framework. Earlier analyses of Van der Waals black holes typically relied on classical thermodynamics or included only limited forms of correction, such as logarithmic terms. Similarly, recent efforts in topological classification have often excluded quantum-corrected geometries. By bridging these gaps, the present study extends the scope of both thermodynamic and topological methodologies, offering a more universal and adaptable classification scheme—one that remains meaningful even in the presence of quantum gravitational effects. In doing so, the study not only deepens our understanding of black hole thermodynamics but also provides a versatile toolkit for future investigations into the quantum structure of spacetime. It lays the groundwork for exploring how microscopic corrections manifest in macroscopic observables and offers a promising path toward reconciling gravity with quantum theory through the lens of thermodynamic geometry and topology.

\section{Acknowledgment}

A.A. is financially supported by the Institute's postdoctoral fellowship at IITK and is grateful to Aditya Singh for the helpful discussions.


\bibliographystyle{unsrt}  
\bibliography{References.bib}

\end{document}